# Leader-Follower Identification Methodology for Non-Lane Disciplined Heterogeneous Traffic Using Steady State Features

**ABSTRACT**

Road traffic in developing countries, such as India, features a heterogeneous mix of vehicles operating under weak lane discipline (HWLD), encompassing both motorised and non-motorised modes with diverse sizes and manoeuvrability. These conditions lead to complex driver interactions, complicating the reliable identification of vehicle-following (VF) behaviour and leader–follower (LF) pairs. Traditional identification methods based on fixed thresholds for longitudinal and lateral proximity often misclassify non-following instances as valid LF pairs, degrading model performance. This study presents a refined and adaptive method for LF identification in HWLD traffic. It employs vehicle-type- and speed-specific desirable gap thresholds derived from the fundamental density–speed diagram to eliminate false-positive pairs. Additionally, Mexican Hat Wavelet Transform (MWT) is employed to analyse LV and SV speed profiles, verifying LV-SV interaction for LF pair identification. The three-stage filtering includes: (i) speed–gap consistency, (ii) approach/diverge detection via relative velocity sign changes and gap range, and (iii) wavelet-based speed correlation using MWT to confirm LV influence on SV. The framework effectively filters out LF pairs associated with overtaking, tailgating, and inconsistent gap dynamics, retaining only those with consistent VF behaviour and improving model accuracy. Analysis across thirteen LF combinations shows that VF dynamics depend on both SV and LV types. Symmetric pairs (e.g., CAR–CAR, AUTO–CAR) exhibit higher predictability and lower errors, while asymmetric pairs with heavy vehicles or two-wheelers show greater variability. The framework offers a robust foundation for traffic modelling and behaviour analysis.

**INTRODUCTION**

Developed countries typically exhibit homogeneous, lane-based traffic, allowing clearer distinction between vehicle-following (VF) and lane-changing behaviours. In contrast, developing nations have heterogeneous, weak-lane-disciplined (HWLD) traffic, where motorized and non-motorized vehicles of varying sizes and kinematic characteristics share road space, leading to complex interactions. In such environments, drivers navigate both longitudinally and laterally by utilizing gaps, favouring an area-based arrangement to optimize space (*1*).

Analysing VF behaviour in HWLD traffic presents unique challenges, particularly in the accurate identification of leader-follower (LF) vehicle pairs, where the motion of a subject vehicle (SV) is influenced by a leader vehicle (LV). However, LF identification in HWLD traffic remains largely underexplored. Therefore, this section presents the limited number of studies that develop various criteria for identifying LF pairs in HWLD traffic. Studies often rely on synthetic or simulated data (*2, 3*) or aggregated macroscopic measures (*4, 5*) for LF identification that fail to capture nuanced individual interactions.

Longitudinal clearance between the assumed leader and follower is a common method used to define the LV and its influence zone on the SV (*1, 6–8*). Lateral clearance or overlap width is another important factor for identifying the LV (*1, 6–10*). A minimum threshold for the continuous following duration was used in a previous study (*8*). The most influential LV was identified based on the closest gap between the SV and LV, particularly when more than one LV was present (*1, 6*)

In HWLD traffic, the influence of other surrounding vehicles between the leader and follower must be considered. Smaller vehicles, such as two-wheelers (TW), often partially occupy the influence region, causing intermittent VF behaviour. Such dynamics, uncommon in lane-based traffic, frequently occur in non-lane-based conditions due to vehicle size variations. One study used space-time plots to identify other influencing vehicles, but noted that visual analysis becomes impractical for large datasets (*9*). A more recent approach proposed a robust methodology for LF identification, focusing on strong LV–SV interaction, significant lateral overlap, and the absence of intervening vehicles (*11*). They observed that an intermediate vehicle between SV and LV weakens their interaction, leading to classification as non-LF pairs. The above literature on the identification of LF pairs reveals several gaps as follows:

- Fixed thresholds for longitudinal clearance, as proposed in the literature, may not capture realistic interactions, as vehicle influence depends on speed, gap, and vehicle class (TW's generally have shorter look-ahead distances than trucks due to the variations in the kinematics characteristics).
- One approach based on Wiedemann-99 driving regimes, is limited to cars and does not generalise well to HWLD traffic with different combinations of LF vehicle types (*11*).
- A potential LV can influence the SV, as seen in similar speed patterns between the two. However, using this relationship for LF pair identification is yet to be explored in the existing literature.
- Although WTs have been applied to traffic flow analysis, their use for LF identification under HWLD traffic remains unexplored. One study applied Mexican Hat Wavelet Transform (MWT) to analyse speed profiles and detect abrupt changes, but its potential for LF identification has yet to be examined (*12*).

This study addresses these gaps by proposing an enhanced LF identification method that incorporates vehicle-type and speed-specific desirable longitudinal gaps estimation using the density-speed (k–v) fundamental diagram (FD) and dynamic speed profile matching via MWT. By leveraging these criteria, the methodology demonstrates improved accuracy over existing heuristic approaches, with broader applicability to diverse vehicle types in HWLD traffic.

**METHODOLOGY**

This section presents an overview of the proposed LF pair identification methodology using the FDs and MWTs. First, the data used in this study are discussed. Next, the LF pairs identified based on the literature and the proposed methodology are presented.

**Data**

This study utilizes open-source HWLD trajectory data from Saidapet, Chennai, as provided in (*13*). The processed data include individual trajectories of 3005 vehicles, with each vehicle's trajectory including the spatial position, speed, and acceleration/deceleration values in both the longitudinal and lateral directions at a 0.5 s resolution.

**Identify leader-follower pairs based on the literature**

The development of a longitudinal response model begins with identifying LF pairs, a challenging task in HWLD traffic conditions due to varying vehicle dimensions, intermittent following, and multiple potential leaders. The following four criteria, based on existing literature, were employed to identify the base LF pairs from the data:

   i. *Longitudinal threshold*: the leader is expected to be present within 30 m from the front bumper of the SV (*1, 6–8*)
   ii. *Lateral overlap*: LV's lateral dimensions should be fully or partially overlapping with SV (*1, 6–10*).
   iii. *Duplicate leader*: when more than one LV is present, the most influential LV is identified based on the closest gap with the SV (*1, 6*)
   iv. *Minimum duration*: A minimum of 5 seconds of continuous following duration (*8, 11, 14, 15*).

Based on these four criteria from the literature, 2125 LF pairs were identified from the Chennai trajectory data. The class-wise distribution of these pairs is presented in **Table 1**. This study specifically examines 13 different LF pair combinations that have a minimum of 20 LF pairs.

**Table 1 LF pairs identified based on the literature**

| SV Type | # of pairs | LF Pair | # of pairs | # of data points |
|---|---|---|---|---|
| CAR | 707 | TW-CAR | 209 | 3273 |
| | | HV-CAR | 30 | 547 |
| | | LCV-CAR | 19 | 313 |
| | | AUTO-CAR | 86 | 1442 |
| | | CAR-CAR | 363 | 6555 |
| AUTO | 352 | AUTO-AUTO | 57 | 1013 |
| | | CAR-AUTO | 94 | 1442 |
| | | TW-AUTO | 176 | 2791 |
| | | HV-AUTO | 19 | 353 |
| | | LCV-AUTO | 6 | 80 |
| TW | 942 | CAR-TW | 183 | 2989 |
| | | HV-TW | 38 | 628 |
| | | LCV-TW | 9 | 148 |
| | | AUTO-TW | 157 | 2750 |
| | | TW-TW | 555 | 8638 |
| HV | 90 | CAR-HV | 24 | 385 |
| | | TW-HV | 47 | 684 |
| | | AUTO-HV | 11 | 188 |
| | | HV-HV | 5 | 92 |
| | | LCV-HV | 3 | 53 |
| LCV | 34 | CAR-LCV | 12 | 214 |
| | | AUTO-LCV | 4 | 80 |
| | | TW-LCV | 15 | 237 |
| | | HV-LCV | 3 | 53 |
| | | LCV-LCV | 0 | 0 |

**Proposed LF identification methodology**

This study introduces two-stage filtrations for identifying LF pairs in HWLD traffic – Gap threshold-based and speed-correlation-based.

i. **Vehicle-type-specific longitudinal gap thresholds**

The commonly used fixed threshold of 30 meters, as used in the literature, for identifying LVs may not accurately capture real-world interactions. For instance, slower-moving vehicles are less influenced by distant LVs, and the look-ahead distance for TWs is typically shorter than that of heavy vehicles such as trucks. To address this, the study proposes using the equilibrium longitudinal gap for different vehicle types based on their instantaneous speeds.

*Desirable Longitudinal Gap:*

The desirable or equilibrium gap (s) is calculated for each SV type based on speed, using density (k) derived from the fundamental q-k diagram (see Error! Reference source not found. **a**) and calculated through Equations (1) and (2) for each vehicle type. The parameters for the fundamental diagram—such as maximum flow, jam density, critical density, and free-flow speed—are sourced from the literature (*16*). Error! Reference source not found. **(b)** illustrates vehicle-type-specific fundamental diagrams constructed using these parameters.

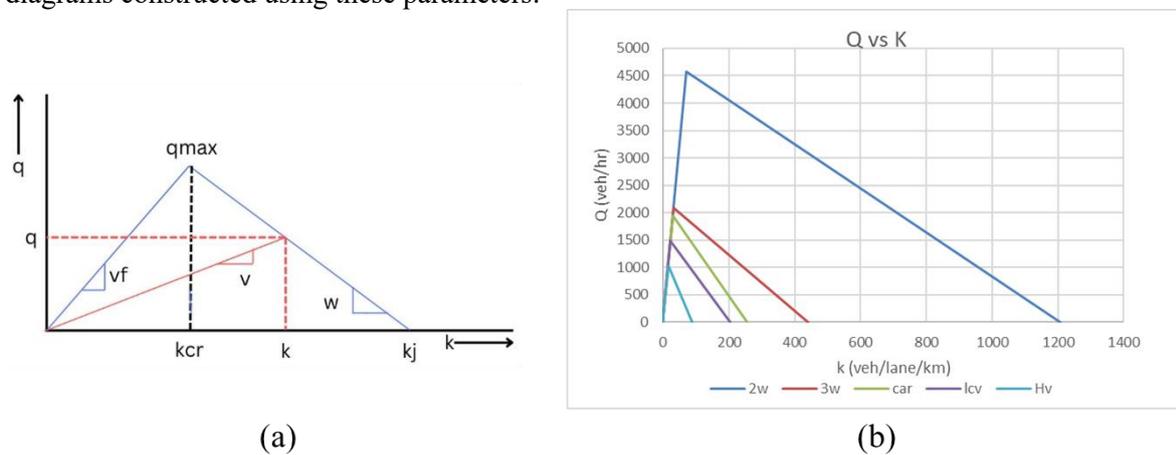

(a)            (b)

**Figure 1 Fundamental diagram (a) homogeneous traffic (b) heterogeneous traffic**

$$k = \frac{w*kj}{(w+v)} \qquad (1)$$

$$s = \frac{1}{k} = \frac{(w+v)}{w*kj} * 1000 \qquad (2)$$

Where s is the equilibrium gap between each class where both leader and follower are of the same class.

As the fundamental diagram parameters for each vehicle class are readily available in existing studies, the q–k relationship is used in this work instead of directly using a k–v (density–speed) relationship. **Table 2** represents the vehicle-specific desirable longitudinal gaps corresponding to different speed values estimated using the above equations based on the FD parameters. The values corresponding to the SV are used irrespective of the LV class. LF pairs are then selected based on whether the observed longitudinal gap falls within the estimated desirable gap range.

**Table 2: Vehicle and speed specific desirable gap from FD**

| Speed (kmph) | Longitudinal spacing (m) threshold | | | | |
|---|---|---|---|---|---|
| | TW | CAR | HV | LCV | AUTO |
| 5 | 1.86 | 6.08 | 15.08 | 7.97 | 4.54 |
| 10 | 2.90 | 8.25 | 19.05 | 11.03 | 6.80 |
| 15 | 3.93 | 10.42 | 23.02 | 14.09 | 9.07 |
| 20 | 4.97 | 12.59 | 26.98 | 17.16 | 11.34 |
| 25 | 6.00 | 14.76 | 30.95 | 20.22 | 13.61 |
| 30 | 7.04 | 16.93 | 34.92 | 23.28 | 15.87 |
| 35 | 8.07 | 19.10 | 38.89 | 26.35 | 18.14 |
| 40 | 9.11 | 21.27 | 42.86 | 29.41 | 20.41 |
| 45 | 10.14 | 23.44 | 46.83 | 32.48 | 22.68 |
| 50 | 11.18 | 25.61 | 50.79 | 35.54 | 24.94 |
| 55 | 12.21 | 27.78 | 54.76 | 38.60 | 27.21 |
| 60 | 13.25 | 29.95 | 58.73 | 41.67 | 29.48 |
| 65 | 14.28 | 32.12 | 62.70 | 44.73 | 31.75 |

  ii. **Speed profile correlation between LV and SV**

Existing methods primarily identify LVs based on relative position without confirming their influence on the SV. The following steps address this issue by evaluating the correlation between the speed profiles of LV and SV using the MWT.

*Speed Profile Correlation:*
Mexican hat wavelet energy trends were analysed to identify LVs that influence SV behaviour. LF pairs were validated if the SV's wavelet energy profile exhibited a lagged similarity to the LV's profile. For LF pairs identified through the prior modification, the LV's influence was assessed by comparing MWT energy plots of LV and SV speeds. LF pairs with at least one matching peak point within a lag of 2 seconds in their wavelet energy plots were shortlisted. **Figure 2** illustrates the wavelet energy profiles of LF pair 408-410.

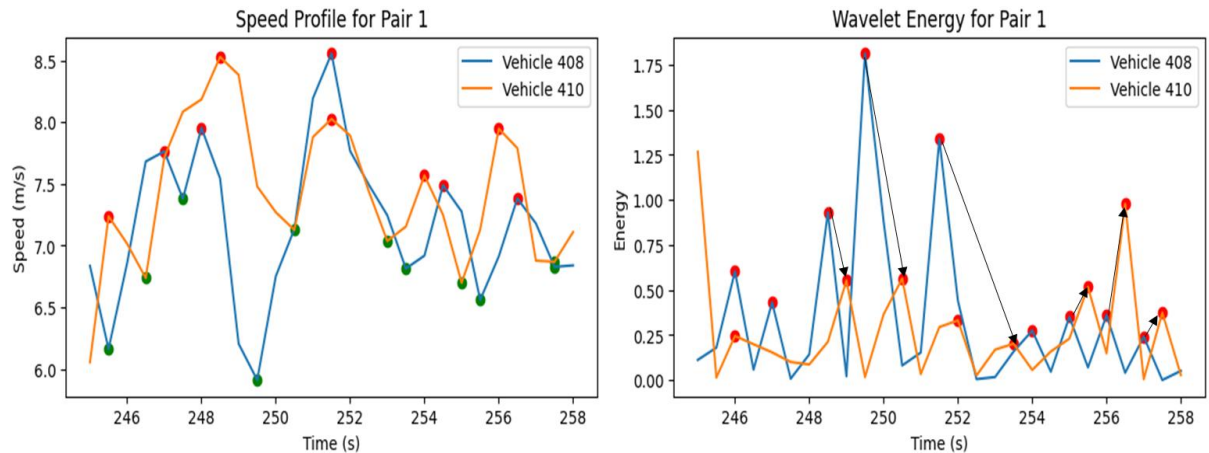

**Figure 2 Sample LF pair (a) speed profile (b) wavelet energy**

Based on the above two stages, four approaches were used to identify the probable outlier pairs. Details of filtered pairs in each of the approaches are given below the figures (Figures 7-10).

**Filtering Steps**

First, the speed and longitudinal gap behaviour are analysed to understand the existing trends among the identified LF pairs shortlisted based on the methods from the literature and presented in Table 1. For demonstrating the steps, CAR-CAR pair was considered in the rest of the document. However, all combinations of LF pairs are analysed in this study, and the results are presented in the subsequent sections. Figure 3 represents the range of speed and longitudinal gap of CAR-CAR pairs. For CAR-CAR, 33 kmph and 11 kmph were the upper and lower whisker points for speed, and 20 m and 0 m were the corresponding points for longitudinal gap.

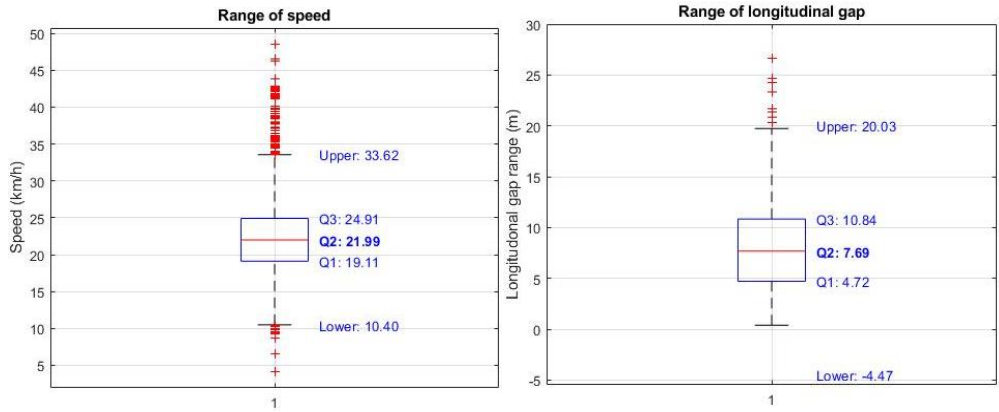

**Figure 3 Box plot of a) speed and b) longitudinal gap**

The relationship between vehicle speed and longitudinal gap, critical for both safety and efficiency, is considered in this step. Figure 4 (a) and Figure 4 (b) gives the speed distribution trends across the gap bins and gap distribution trends across the speed bins, respectively.

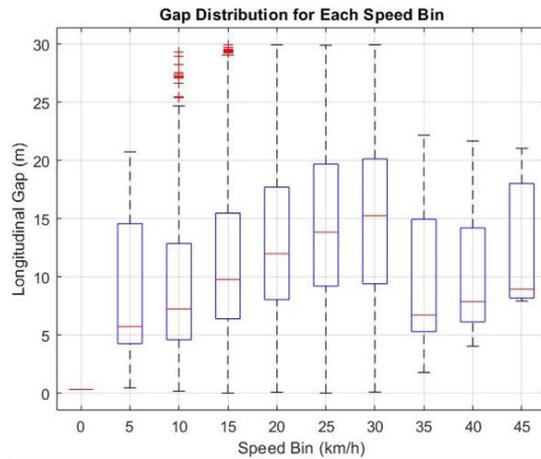

(a)

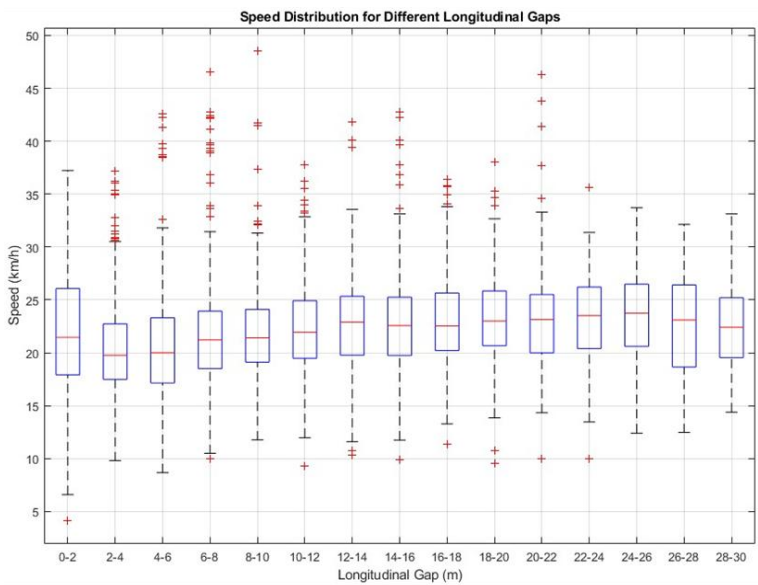

(b)

**Figure 4 Distribution plot (a) gap distribution across speed bin, (b) speed distribution across gap bin**

Generally, the speed of the SV increases when the longitudinal gap to the LV increases, exhibiting a positive correlation between the two. This is because drivers require more space to react and brake safely at higher speeds. However, the above plots do not exhibit an increasing trend for longitudinal gap vs longitudinal speed.

Using the base pairs, **Figure 4 (a)** shows that there is a difference in behaviour when the speed is greater than 35 km/h in the speed distribution plot. This demonstrates a transition stage. This might be the case where SV tries to overtake the LV, thus exhibiting higher speed and lower longitudinal gap. Similarly, the trend is different for the first and last gap bins in gap distribution plots **(Figure 4 b)**. At smaller gap bins, large speed values are coming; this might be the transition stages where SV exhibit a tail gate tendency. Also, there are cases where SV shifts laterally, thereby making a shift from strictly following to partially overlapping or non-overlapping following behaviour. Under such conditions, SV maintains a lesser longitudinal gap at higher speeds.

This might be because not all the identified pairs are in a vehicle-following behaviour. Although the SV appears to be physically positioned behind the LV with lateral overlap, its behaviour might not be under the influence of the LV. Thus, each pair needs to be examined in detail to identify the underlying behaviour.

As an initial step, we identify the probable outlier pair based on the gap bin and speed bin plots using the following criteria:
  i. Longitudinal gaps less than the 5th percentile within each speed bin.
  ii. Longitudinal gaps greater than the 95th percentile within each speed bin.
  iii. Longitudinal speed exceeding the upper whisker point (33 km/h for CAR-CAR).
  iv. Longitudinal speed below the lower whisker point (11 km/h for CAR-CAR).

Then, we examine each of the potential outlier pairs graphically using the following plots:
  i. Longitudinal and lateral trajectories vs time.
  ii. Speed versus longitudinal gap scatter plots (with sequentially numbered points).
  iii. Hysteresis plot (relative longitudinal velocity vs. longitudinal gap (*17*).
  iv. Oblique plot of modified longitudinal position versus time (*18*)
  v. Lateral gap versus lateral speed.
  vi. Lateral gap vs. time and lateral speed vs. time.
  vii. Box plots of relative longitudinal velocity, longitudinal gap, longitudinal speed, lateral gap, and lateral speed.

Based on the trends, such as similarities/ dissimilarities of trajectory plots (both actual and oblique plot), hysteresis/ non-hysteresis behaviour of gap vs relative velocity (both longitudinal and lateral) and outliers of box plots (of relative velocity, longitudinal gap, longitudinal gap, longitudinal speed, lateral gap and lateral speed) pairs can be identified as following/ non-following.

**Figure 5** represents an anomalous behaviour pattern for pair 1758-1761. Here, the SV appears to be following behaviour, but actually, it is shifting laterally, trying for an opportunity to overtake the LV. However, this tendency is not captured if only the hysteresis plot and the oblique plot are considered. This showcases the need for analysis based on both longitudinal and lateral behaviour.

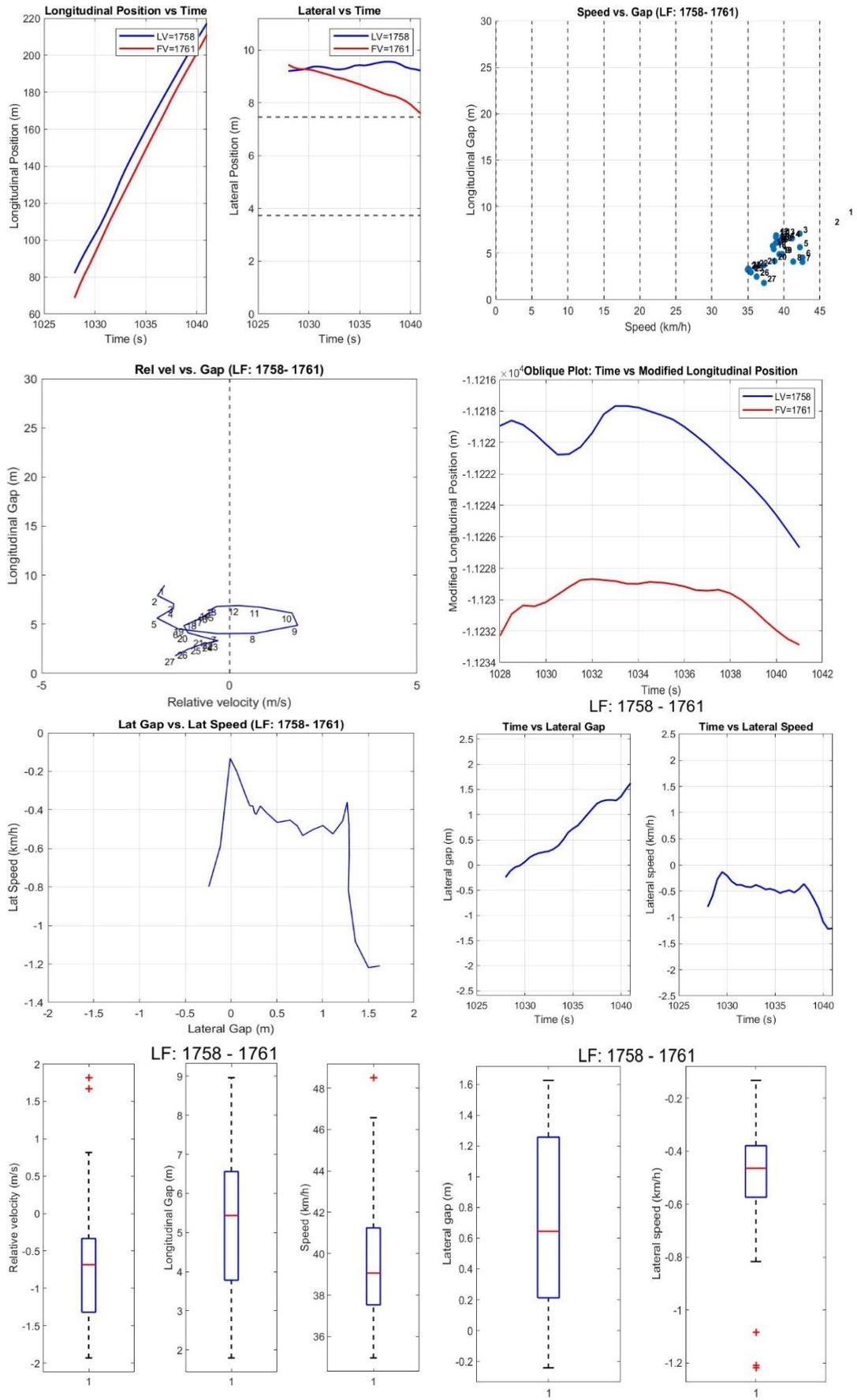

**Figure 5 LF pair 1758-1761**

However, it is not practically possible to analyse each of the pairs based on these different types of graphical representation. Therefore, we propose the optimal threshold values for relative velocity, lateral gap, and range of longitudinal gap to filter out non-influencing LVs.

On further detailed analysis of those cases, the following list of thresholds was observed from the pairs that lack the vehicle following behaviour for the study dataset. However, these thresholds need to be recalibrated for other datasets.

i. Relative velocity greater than 2.5 m/s and less than -2.5 m/s.
ii. Lateral gap greater than 1.5m and less than -1.5m. If the SV moves towards the right side of LV, it is taken as positive and negative for the left side movement.
iii. Maintaining a longitudinal gap of less than 2m at higher speeds
iv. Maintaining a large longitudinal gap greater than 28m at moderate speeds.
v. Relative velocity sign change ratio <0.3.

In typical car-following scenarios, the relationship between relative velocity and longitudinal gap exhibits hysteresis behaviour. However, when vehicles are either diverging or converging without actual following interaction, this hysteresis is absent. For instance, in an approaching scenario, such as LF pair 1823–1830 (refer to Figure 6), the relative velocity remains consistently positive, indicating a lack of hysteresis and the longitudinal gap range is very large. To quantify this behaviour, two additional conditions are introduced as mentioned below

i. Relative velocity sign change ratio, $r$, is calculated as:

$$r = \frac{Number\ of\ sign\ changes\ in\ relative\ velocity\ of\ a\ pair}{Total\ number\ of\ data\ points\ in\ a\ pair}$$

Higher sign change ratio reflects the dynamic switching between acceleration and deceleration behaviours, which is typically present in genuine car-following pairs. For a proper vehicle following behaviour pair, there will be good hysteresis behaviour, and the sign change ratio will be high. The max possible value is 0.5. It was found that probable pairs with a sign change ratio < 0.3 showed significant outlier behaviour.

ii. Longitudinal gap range, calculated as

$$t^M = \sup\ (x_i(t) - x_{i+1}(t)) \qquad (3)$$

$$t^m = \inf\ (x_i(t) - x_{i+1}(t)) \qquad (4)$$

$$range = \left|\left(x_i(t^M) - x_{i+1}(t^M)\right) - \left(x_i(t^m) - x_{i+1}(t^m)\right)\right| \qquad (5)$$

where, $x_i(t)$ and $x_{i+1}(t)$ are the positions of LV and SV at time t respectively. $t^M$ and $t^m$ are the times at which the longitudinal gaps are maximum and minimum.

These two filters take care of the converging and diverging cases. The current study used a threshold value of 0.3 and 10m for relative velocity ratio and longitudinal gap.

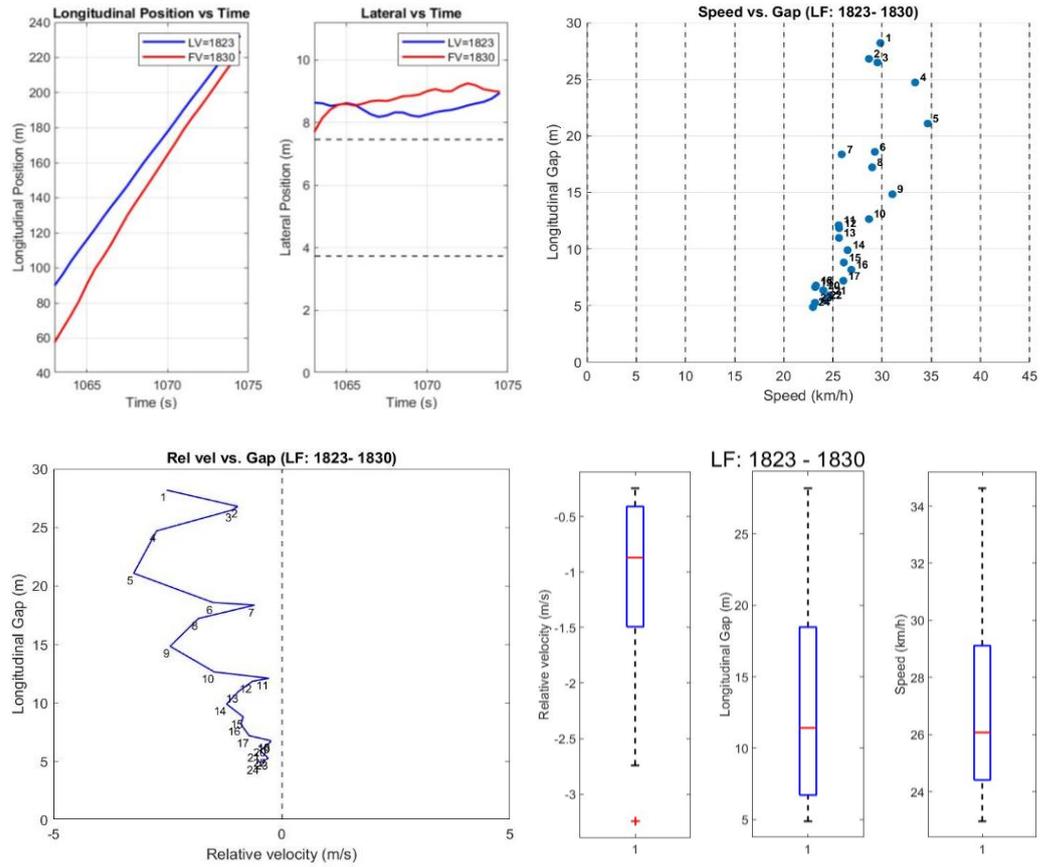

**Figure 6 LF: 1823-1830 a) Trajectory b) Scatter plot c) Hysteresis plot and d) Box plots**

Finally, the retained pairs, are checked for the speed correlation between them. The Mexican hat wavelet energy plots of the speed profiles are compared for delayed similarities to identify influencing LVs.

All the filtering conditions discussed in this section are grouped into four approaches and implemented on the study dataset, as explained in the next subsection. The best approach is then selected based on the model's performance.

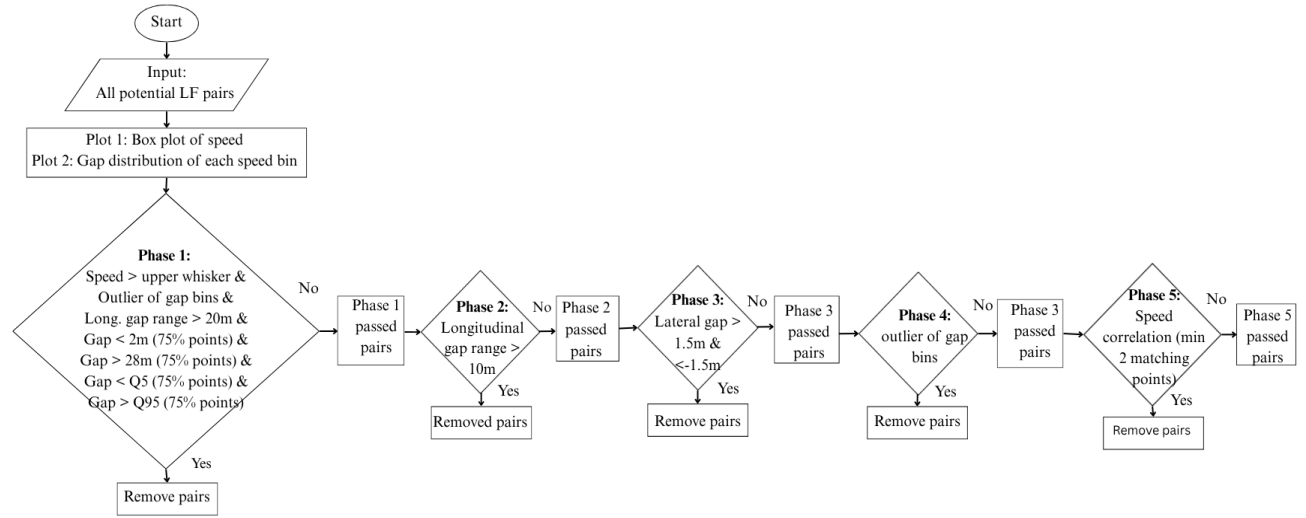

**Figure 7 Filtering outlier pairs- Approach 1**

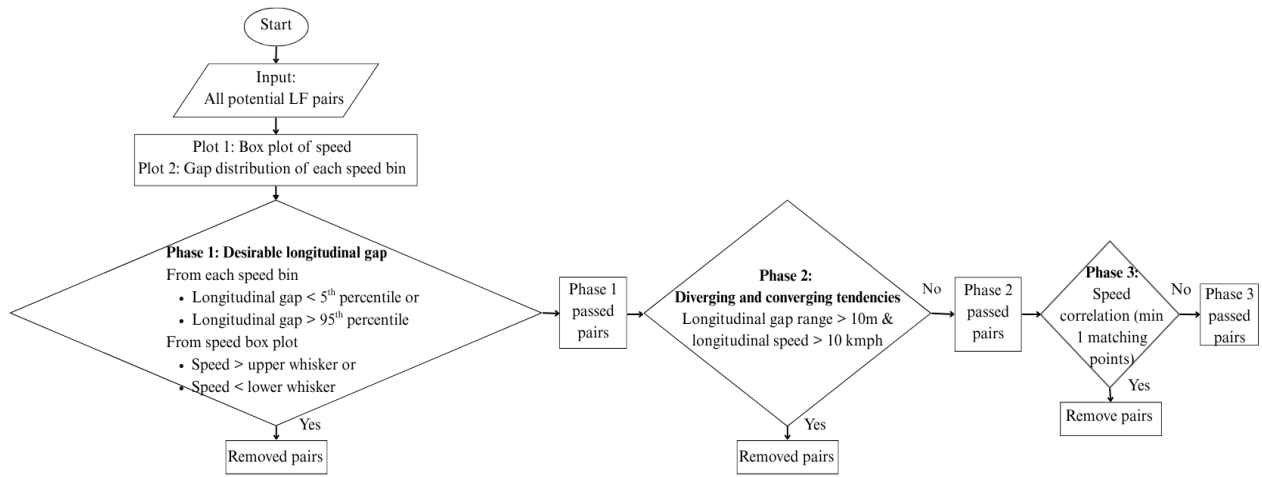

Figure 8 Filtering outlier pairs- Approach 2

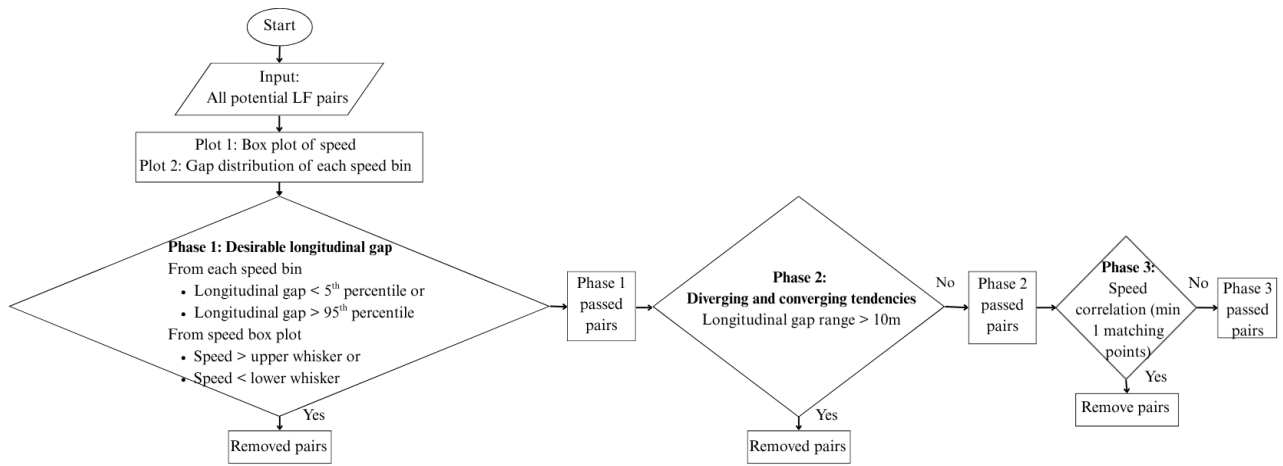

Figure 9 Filtering outlier pairs- Approach 3

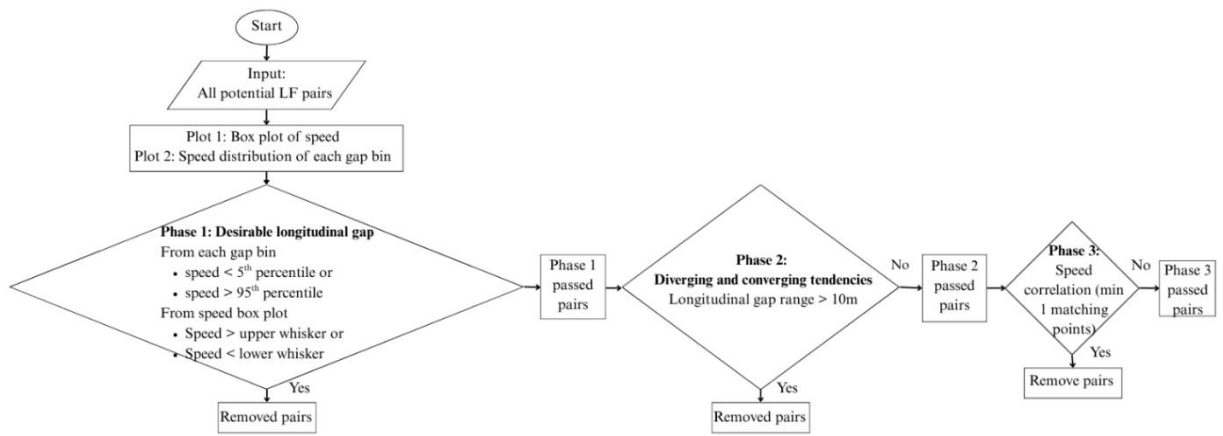

Figure 10 Filtering outlier pairs- Approach 4

Among these, Approach 4 produced the most promising results (as discussed in the following section). Using this method, a total of 167 LF pairs were shortlisted and retained for subsequent modelling. The proposed methodology is summarised as the following algorithm for ease of implementation for any new dataset:

---

**Algorithm: Filtering Non-Following Leader-Follower (LF) Pairs**

**Step 1**: Identify all vehicle-specific LF vehicle pairs (e.g. CAR-CAR) from trajectory data based on literature.

**Step 2**: Construct longitudinal gap bin plots using an appropriate bin size, based on the desirable gap range (e.g., 2 meters for CAR–CAR, derived from the fundamental diagram). For each bin, generate box plots of the follower vehicle's speed.

**Step 3**: Detect potential outliers in the longitudinal gap data using the following conditions:

  i. Longitudinal gaps less than the 5th percentile within each speed bin.
  ii. Longitudinal gaps greater than the 95th percentile within each speed bin.
  iii. Longitudinal speed exceeding the upper whisker (i.e. 33 km/h).
  iv. Longitudinal speed below the lower whisker (i.e. 11 km/h).

**Step 4:** Apply the following sub criteria to remove outlier pairs:

  i. Relative velocity exceeds ±2.5 m/s.
  ii. Lateral gap exceeds ±1.5 m (positive: SV is to the right of LV; negative: SV is to the left).

**Step 5**: Visually inspect the identified outlier LF pairs using the following plots.

  i. Longitudinal and lateral trajectories vs time.
  ii. Speed versus longitudinal gap scatter plots (with sequentially numbered of points).
  iii. Hysteresis plot (relative longitudinal velocity vs. longitudinal gap).
  iv. Oblique plot of modified longitudinal position versus time.
  v. Lateral gap versus lateral speed.
  vi. Lateral gap vs. time and lateral speed vs. time.
  vii. Box plots of relative longitudinal velocity, longitudinal gap, longitudinal speed, lateral gap, and lateral speed.

Based on below conditions confirm non-following behaviour and remove the identified outlier LF pairs

  i. Mismatching patterns in the longitudinal, lateral, and oblique plots.
  ii. Skewed hysteresis behaviour in the relative longitudinal velocity vs. gap plot.
  iii. Absence of hysteresis behaviour in the lateral gap vs. lateral speed plot (i.e., distributed behaviour).
  iv. Isolated outlier points in the speed vs. longitudinal gap scatter plot.

If an outlier point is located at the start or end of the dataset, remove it, along with any consecutive points preceding or following it. If the outlier appears in the middle of valid data, retain it to preserve the consistency of the time series.

**Step 8**: To the cleaned dataset, apply the second set of condition to filter out the diverging and converging behaviour. Filter out pairs that satisfy both the two conditions given below.

  i. Longitudinal gap range (max gap – min gap) for each pair exceeding 10 meters.
  ii. Relative velocity sign change ratio less than 0.3.

**Step 9**: Repeat **Step 5** to validate and remove the additional outlier pairs and points visually.

**Step 10**: For the final set of LF pairs, assess the influence of the LV by comparing the Mexican hat wavelet energy plots of the LV and the SV speeds. Retain only those LF pairs that exhibit at least one matching peak in their energy plots (selected based on MATLAB code).

---

**Performance Evaluation**

In addition to conventional performance metrics such as R², Adjusted R², MAE, and RMSE, the current study uses two supplementary indicators to evaluate the effectiveness of the proposed LF pair filtering methodology:

    i.   Normalized RMSE (NRMSE)

In order to provide a scale-independent measure of model performance, the Normalized Root Mean Square Error (NRMSE) is introduced. This metric normalizes the RMSE by the standard deviation of the observed dependent variable (SV acceleration), thereby facilitating consistent performance comparisons across different LF pair categories.

$$NRMSE = \frac{RMSE}{\sigma_Y} \qquad (6)$$

    ii.   Distribution of weights

Weights for each data point were computed based on the concept of Weighted Linear Regression (WLR). Initially, an Ordinary Least Squares (OLS) regression model was fitted to the complete dataset to estimate residuals. The weights were then calculated as the inverse of the absolute residuals as shown in Equations (7) and (8):

$$r_i = y_i - \hat{y}_i \qquad (7)$$

where, $y_i$ is the observed acceleration value and $\hat{y}_i$ is the predicted acceleration value from OLS.

$$w_i = \frac{1}{|r_i| + \varepsilon}, \varepsilon = 10^{-6} \qquad (8)$$

A higher weight corresponds to a smaller residual, implying greater significance of the corresponding data point in the regression model. The weight distribution of both retained and removed pairs was estimated for each of the LF pair combinations.

The effectiveness of the proposed methodology is evaluated against existing approaches in the subsequent results section.

**RESULTS AND DISCUSSION**

    This section evaluates the performance of the existing LF identification method against the proposed approach. For the performance evaluation, this study aims to predict the longitudinal response (acceleration or deceleration) of the SV using a multiple linear regression model. Three fundamental stimuli are considered for modelling: the relative velocity between the SV and the LV, the longitudinal gap between the SV and the LV, and the SV speed. An earlier study reported optimal acceleration response prediction using a 0.5-second update time with the same open-source Chennai data (*7*). Consequently, this study also adopts a 0.5-second update time. The model structure is represented in Equation 9.

$$y(t + \tau) = \beta_0 + \beta_1 x1(t) + \beta_2 x2(t) + \beta_3 x3(t) + \varepsilon \qquad 9)$$

    Where $y(t + \tau)$ is the acceleration or deceleration response of the SV at time $(t + \tau)$, $x1$ is the relative speed between LV and SV, $x2$ is the bumper-to-bumper gap between LV and SV in the longitudinal direction, $x3$ is the SV speed, $\beta_x$ is the parameter associated with variable $x$ and $\varepsilon$ is the error term.

For demonstration, CAR–CAR pairs are used to evaluate all approaches. The modelling results and performance comparisons are summarised in Tables 3-6.

**Table 3 Performance matrix -Approach 1**

| Approach 1 | | | | | |
|---|---|---|---|---|---|
| CAR-CAR | | $R^2$ | Adj $R^2$ | MAE | RMSE |
| Pairs based on literature (363 pairs) | Train set | 0.250 | 0.249 | 0.548 | 0.778 |
| | Test set | 0.275 | 0.274 | 0.546 | 0.765 |
| Iteration 1 (303 pairs) | Train set | 0.294 | 0.293 | 0.488 | 0.652 |
| | Test set | 0.326 | 0.324 | 0.487 | 0.649 |
| Iteration 2 (262 pairs) | Train set | 0.341 | 0.340 | 0.496 | 0.660 |
| | Test set | 0.365 | 0.364 | 0.460 | 0.624 |
| Iteration 3 (221 pairs) | Train set | 0.381 | 0.380 | 0.478 | 0.635 |
| | Test set | 0.348 | 0.347 | 0.488 | 0.652 |
| Iteration 4 (213 pairs) | Train set | 0.369 | 0.368 | 0.474 | 0.630 |
| | Test set | 0.386 | 0.384 | 0.477 | 0.636 |
| Iteration 5 min 3 matching points (80 pairs) | Train set | 0.369 | 0.368 | 0.474 | 0.630 |
| | Test set | 0.386 | 0.384 | 0.477 | 0.636 |
| Iteration 5 min 2 matching points (130 pairs) | Train set | 0.386 | 0.385 | 0.485 | 0.648 |
| | Test set | 0.353 | 0.351 | 0.487 | 0.653 |

**Table 4 Performance matrix -Approach 2**

| Approach 2 | | | | | |
|---|---|---|---|---|---|
| CAR-CAR | | $R^2$ | Adj $R^2$ | MAE | RMSE |
| Pairs based on literature (363 pairs) | Train set | 0.250 | 0.249 | 0.548 | 0.778 |
| | Test set | 0.275 | 0.274 | 0.546 | 0.765 |
| Iteration 1 (266 pairs) | Train set | 0.325 | 0.325 | 0.478 | 0.635 |
| | Test set | 0.308 | 0.307 | 0.476 | 0.626 |
| Iteration 2 (207 pairs) | Train set | 0.335 | 0.334 | 0.452 | 0.600 |
| | Test set | 0.334 | 0.332 | 0.445 | 0.593 |
| Iteration 3 (172 pairs) | Train set | 0.325 | 0.325 | 0.445 | 0.587 |
| | Test set | 0.331 | 0.329 | 0.448 | 0.601 |

**Table 5 Performance matrix -Approach 3**

| Approach 3 | | | | | |
|---|---|---|---|---|---|
| CAR-CAR | | $R^2$ | Adj $R^2$ | MAE | RMSE |
| Pairs based on literature (363 pairs) | Train set | 0.250 | 0.249 | 0.548 | 0.778 |
| | Test set | 0.275 | 0.274 | 0.546 | 0.765 |
| Iteration 1 (266 pairs) | Train set | 0.325 | 0.325 | 0.478 | 0.635 |
| | Test set | 0.308 | 0.307 | 0.476 | 0.626 |
| Iteration 2 (251 pairs) | Train set | 0.327 | 0.326 | 0.480 | 0.632 |
| | Test set | 0.342 | 0.341 | 0.471 | 0.635 |
| Iteration 3 (194 pairs) | Train set | 0.330 | 0.329 | 0.462 | 0.619 |
| | Test set | 0.309 | 0.307 | 0.484 | 0.629 |

**Table 6 Performance matrix -Approach 4**

| Approach 4 | | | | | |
|---|---|---|---|---|---|
| CAR-CAR | | $R^2$ | Adj $R^2$ | MSE | RMSE |
| Pairs based on literature (363 pairs) | Train set | 0.250 | 0.249 | 0.605 | 0.778 |
| | Test set | 0.275 | 0.274 | 0.585 | 0.765 |
| Iteration 1 (225 pairs) | Train set | 0.329 | 0.329 | 0.359 | 0.599 |
| | Test set | 0.331 | 0.329 | 0.350 | 0.591 |

|  |  |  | Train set | 0.361 | 0.361 | 0.361 | 0.601 |
| --- | --- | --- | --- | --- | --- | --- | --- |
| Iteration 2 (206 pairs) | | | Test set | 0.292 | 0.290 | 0.351 | 0.593 |
| Iteration 3 (167 pairs) | | | Train set | 0.342 | 0.341 | 0.360 | 0.600 |
| | | | Test set | 0.336 | 0.334 | 0.340 | 0.583 |

Among the four approaches evaluated, Approach 4 demonstrated the best performance. Using this method, 167 CAR–CAR pairs were shortlisted and used for subsequent modelling. The same filtering methodology was applied to the remaining twelve LF pair combinations.

The effectiveness of the proposed methodology was analysed based on the following criteria

**Performance matrix**

To assess model performance, K-fold cross-validation was employed for model development, with five folds used to ensure robustness and minimise overfitting. Due to space limitations, only the average performance metrics across the five folds are reported for each LF pair combination (Refer Table 7).

**Table 7 Performance matrix of proposed filtering methodology across LF pairs**

| SV | LF pair | Sample | | R2 | Adj R2 | MAE | RMSE | NRMSE |
| --- | --- | --- | --- | --- | --- | --- | --- | --- |
| CAR | CAR-CAR | Before filtering | Train | 0.258 | 0.257 | 0.548 | 0.774 | 0.862 |
| | | | Test | 0.256 | 0.254 | 0.548 | 0.774 | 0.862 |
| | | After filtering | Train | 0.341 | 0.340 | 0.448 | 0.595 | 0.812 |
| | | | Test | 0.338 | 0.335 | 0.449 | 0.595 | 0.814 |
| | TW-CAR | Before filtering | Train | 0.214 | 0.213 | 0.527 | 0.725 | 0.887 |
| | | | Test | 0.203 | 0.199 | 0.528 | 0.728 | 0.892 |
| | | After filtering | Train | 0.296 | 0.295 | 0.459 | 0.627 | 0.839 |
| | | | Test | 0.283 | 0.278 | 0.461 | 0.630 | 0.846 |
| | HV-CAR | Before filtering | Train | 0.389 | 0.385 | 0.550 | 0.748 | 0.781 |
| | | | Test | 0.345 | 0.327 | 0.557 | 0.756 | 0.807 |
| | | After filtering | Train | 0.382 | 0.375 | 0.508 | 0.650 | 0.786 |
| | | | Test | 0.355 | 0.325 | 0.513 | 0.658 | 0.802 |
| | AUTO-CAR | Before filtering | Train | 0.366 | 0.364 | 0.527 | 0.716 | 0.797 |
| | | | Test | 0.352 | 0.346 | 0.528 | 0.719 | 0.804 |
| | | After filtering | Train | 0.391 | 0.389 | 0.462 | 0.613 | 0.780 |
| | | | Test | 0.371 | 0.361 | 0.465 | 0.617 | 0.792 |
| TW | CAR-TW | Before filtering | Train | 0.291 | 0.290 | 0.595 | 0.799 | 0.842 |
| | | | Test | 0.286 | 0.282 | 0.596 | 0.799 | 0.845 |
| | | After filtering | Train | 0.365 | 0.364 | 0.529 | 0.699 | 0.797 |
| | | | Test | 0.360 | 0.355 | 0.530 | 0.700 | 0.800 |
| | TW-TW | Before filtering | Train | 0.297 | 0.297 | 0.621 | 0.818 | 0.838 |
| | | | Test | 0.295 | 0.294 | 0.622 | 0.819 | 0.840 |
| | | After filtering | Train | 0.332 | 0.332 | 0.578 | 0.762 | 0.817 |
| | | | Test | 0.330 | 0.327 | 0.579 | 0.763 | 0.819 |
| | HV-TW | Before filtering | Train | 0.346 | 0.342 | 0.628 | 0.829 | 0.809 |
| | | | Test | 0.325 | 0.308 | 0.635 | 0.833 | 0.821 |
| | | After filtering | Train | 0.386 | 0.380 | 0.565 | 0.740 | 0.784 |
| | | | Test | 0.358 | 0.331 | 0.572 | 0.746 | 0.801 |
| | AUTO-TW | Before filtering | Train | 0.381 | 0.380 | 0.598 | 0.790 | 0.787 |
| | | | Test | 0.378 | 0.374 | 0.599 | 0.790 | 0.788 |

|  |  |  | Train | 0.402 | 0.400 | 0.539 | 0.720 | 0.773 |
|---|---|---|---|---|---|---|---|---|
|  |  | After filtering | Test | 0.394 | 0.387 | 0.540 | 0.722 | 0.777 |
| AUTO | CAR-AUTO | Before filtering | Train | 0.349 | 0.347 | 0.575 | 0.763 | 0.807 |
|  |  |  | Test | 0.341 | 0.334 | 0.577 | 0.764 | 0.812 |
|  |  | After filtering | Train | 0.377 | 0.374 | 0.527 | 0.719 | 0.789 |
|  |  |  | Test | 0.362 | 0.349 | 0.531 | 0.722 | 0.797 |
|  | TW-AUTO | Before filtering | Train | 0.308 | 0.307 | 0.602 | 0.808 | 0.832 |
|  |  |  | Test | 0.303 | 0.299 | 0.604 | 0.809 | 0.835 |
|  |  | After filtering | Train | 0.361 | 0.359 | 0.553 | 0.727 | 0.799 |
|  |  |  | Test | 0.346 | 0.337 | 0.556 | 0.730 | 0.808 |
|  | AUTO-AUTO | Before filtering | Train | 0.338 | 0.336 | 0.597 | 0.808 | 0.814 |
|  |  |  | Test | 0.318 | 0.308 | 0.606 | 0.817 | 0.825 |
|  |  | After filtering | Train | 0.383 | 0.378 | 0.565 | 0.759 | 0.786 |
|  |  |  | Test | 0.373 | 0.354 | 0.566 | 0.761 | 0.791 |
| HV | CAR-HV | Before filtering | Train | 0.368 | 0.361 | 0.532 | 0.686 | 0.795 |
|  |  |  | Test | 0.338 | 0.311 | 0.537 | 0.688 | 0.811 |
|  |  | After filtering | Train | 0.472 | 0.462 | 0.472 | 0.635 | 0.726 |
|  |  |  | Test | 0.393 | 0.342 | 0.491 | 0.660 | 0.776 |
|  | TW-HV | Before filtering | Train | 0.252 | 0.248 | 0.588 | 0.794 | 0.865 |
|  |  |  | Test | 0.236 | 0.219 | 0.594 | 0.799 | 0.873 |
|  |  | After filtering | Train | 0.361 | 0.354 | 0.499 | 0.680 | 0.799 |
|  |  |  | Test | 0.353 | 0.323 | 0.502 | 0.681 | 0.804 |

In addition, the percentage improvement achieved after applying the filtering methodology is calculated and presented for each LF pair category (refer to Table 8). This allows for a clear comparison of model performance before and after filtering, highlighting the effectiveness of the proposed approach in enhancing the quality of the dataset and improving regression outcomes.

**Table 8 Percentage improvement in the performance matrix**

| SV | LF pair | Improvement (%) | | | | | Outliers Removed (%) |
|---|---|---|---|---|---|---|---|
|  |  | $R^2$ | Adj $R^2$ | MAE | RMSE | NRMSE |  |
| CAR | CAR-CAR | 31.1 | 30.1 | -18.1 | -23.1 | -5.6 | 51.93 |
|  | TW-CAR | 39.6 | 39.2 | -12.7 | -13.4 | -5.2 | 44.24 |
|  | HV-CAR | 2.9 | -0.7 | -7.8 | -13.0 | -0.7 | 38.76 |
|  | AUTO-CAR | 5.3 | 4.3 | -11.9 | -14.1 | -1.5 | 37.73 |
| TW | CAR-TW | 25.9 | 25.7 | -11.2 | -12.4 | -5.3 | 38.41 |
|  | TW-TW | 11.8 | 11.5 | -6.9 | -6.8 | -2.5 | 46.06 |
|  | HV-TW | 10.1 | 7.4 | -9.9 | -10.4 | -2.5 | 38.85 |
|  | AUTO-TW | 4.2 | 3.5 | -9.8 | -8.6 | -1.4 | 45.05 |
| AUTO | CAR-AUTO | 6.0 | 4.4 | -8.0 | -5.6 | -1.7 | 47.02 |
|  | TW-AUTO | 14.3 | 13.0 | -8.0 | -9.8 | -3.2 | 56.72 |
|  | AUTO-AUTO | 17.2 | 14.9 | -6.6 | -6.9 | -4.1 | 49.26 |
| HV | CAR-HV | 16.3 | 10.2 | -8.6 | -4.0 | -4.3 | 48.05 |
|  | TW-HV | 49.3 | 47.6 | -15.5 | -14.8 | -8.0 | 48.83 |

In the case of CAR, the most significant improvements were observed for the CAR-CAR and TW-CAR pairs, followed by the AUTO-CAR and HV-CAR pairs. This trend is consistent with the removal of pairs across different LV types. The vehicle size asymmetry approach proposed in (*1*) is used to categorize pairs into symmetric, positive asymmetric, and negative asymmetric groups. Negative asymmetric pairs (AUTO-CAR and TW-CAR) exhibited smaller improvements compared to symmetric pairs (CAR-CAR). This may be attributed to the high manoeuvrability of LVs, which complicates tracking for larger SVs. Conversely, SVs are less concerned with smaller LVs and may instead be influenced by the next larger LV positioned ahead or nearby. Regarding positive asymmetric pairs, i.e. HV-CAR, the original dataset is of sufficiently high quality with minimal outliers, resulting in comparatively lesser improvements.

In the case of TW, the most significant improvements were observed in CAR-TW and HV-TW, followed by TW-TW and AUTO-TW. Positive asymmetric LF pairs exhibited comparatively better performance and fewer outliers. Although the TW-TW combination had the highest number of outliers removed, its final performance was slightly inferior to that of the other combinations. For AUTO, the highest improvements were observed for AUTO-AUTO, followed by TW-AUTO and CAR-AUTO. Similarly, the highest improvement was observed for the TW-HV case, followed by the CAR-HV case in the HV category.

**Model coefficients**

Based on the results from K-fold cross-validation, the folds corresponding to the best and worst model performance—identified using $R^2$ and Normalized RMSE (NRMSE) values—were selected for detailed coefficient comparison (Refer Table 9).

For the CAR–CAR category, the longitudinal gap became a significant predictor after filtering, indicating that the following behaviour in such pairs is primarily influenced by the longitudinal gap and the subject vehicle (SV) speed, rather than relative velocity. In contrast, for TW–CAR and AUTO–CAR pairs, the significance of longitudinal gap and SV speed decreased, while the influence of relative velocity increased. This suggests that in negative asymmetric combinations involving CARs, the following behaviour is more dependent on relative velocity. A reverse trend is observed for TWs. In positive asymmetric pairs such as CAR–TW and AUTO–TW, the following behaviour is primarily governed by relative velocity, similar to the trend seen in TW–TW combinations. However, HV–TW emerges as an exception, showing less dependency on relative velocity. This could be attributed to the heavy vehicle's slower speed, lower manoeuvrability, and larger size, which may hinder the TW's visibility and force it to follow more cautiously. A trend comparable to that of TWs is also observed for AUTO. In positive asymmetric pairs like CAR–AUTO, vehicle-following behaviour is mainly influenced by relative velocity. In the CAR–HV case, the significance of both relative velocity and longitudinal gap increased, while the importance of SV speed decreased. Similarly, for TW–HV pairs, dependence on both relative velocity and SV speed increased.

**Table 9 Modelling details**

| SV | LF pair | Sample filtering | R² | Adj R² | MAE | RMSE | NRMSE | Variable coefficient | | | | p value | | | |
|---|---|---|---|---|---|---|---|---|---|---|---|---|---|---|---|
| | | | | | | | | Intercept | Relative velocity | Longitudinal gap | SV speed | Intercept | Relative velocity | Longitudinal gap | SV speed |
| CAR | CAR-CAR | Before | 0.273 | 0.272 | 0.552 | 0.748 | 0.853 | 0.014 | 0.320 | 0.004 | -0.209 | 0.181 | 0 | 0.743 | 0 |
| | | After | 0.381 | 0.378 | 0.458 | 0.606 | 0.787 | -0.022 | 0.268 | 0.092 | -0.229 | 0.065 | 0 | 0 | 0 |
| | | Before | 0.236 | 0.234 | 0.540 | 0.783 | 0.874 | 0.016 | 0.312 | 0.017 | -0.221 | 0.135 | 0 | 0.118 | 0 |
| | | After | 0.305 | 0.302 | 0.446 | 0.579 | 0.834 | -0.010 | 0.285 | 0.095 | -0.232 | 0.400 | 0 | 0.000 | 0 |
| | TW-CAR | Before | 0.262 | 0.259 | 0.518 | 0.741 | 0.859 | 0.002 | 0.135 | 0.071 | -0.261 | 0.873 | 0 | 0 | 0 |
| | | After | 0.332 | 0.327 | 0.449 | 0.632 | 0.817 | -0.009 | 0.224 | 0.063 | -0.246 | 0.569 | 0 | 0.0003 | 0 |
| | | Before | 0.121 | 0.117 | 0.557 | 0.755 | 0.937 | -0.012 | 0.163 | 0.098 | -0.294 | 0.394 | 0 | 0 | 0 |
| | | After | 0.210 | 0.203 | 0.431 | 0.596 | 0.889 | 0.004 | 0.210 | 0.093 | -0.290 | 0.817 | 0 | 0 | 0 |
| | HV-CAR | Before | 0.480 | 0.465 | 0.570 | 0.775 | 0.721 | -0.007 | 0.328 | 0.080 | -0.296 | 0.848 | 0 | 0.0274 | 0 |
| | | After | 0.443 | 0.416 | 0.456 | 0.596 | 0.747 | 0.012 | 0.293 | 0.064 | -0.277 | 0.773 | 0 | 0.1225 | 0 |
| | | Before | 0.195 | 0.172 | 0.528 | 0.686 | 0.897 | -0.015 | 0.429 | 0.095 | -0.232 | 0.693 | 0 | 0.0111 | 0 |
| | | After | 0.285 | 0.251 | 0.533 | 0.679 | 0.846 | 0.003 | 0.396 | 0.048 | -0.178 | 0.937 | 0 | 0.2504 | 0.0003 |
| | AUTO-CAR | Before | 0.400 | 0.394 | 0.540 | 0.751 | 0.774 | 0.007 | 0.200 | 0.172 | -0.410 | 0.754 | 0 | 0 | 0 |
| | | After | 0.433 | 0.423 | 0.466 | 0.610 | 0.753 | -0.001 | 0.217 | 0.128 | -0.351 | 0.954 | 0 | 0 | 0 |
| | | Before | 0.287 | 0.280 | 0.493 | 0.681 | 0.844 | -0.023 | 0.175 | 0.190 | -0.466 | 0.291 | 0 | 0 | 0 |
| | | After | 0.270 | 0.257 | 0.462 | 0.597 | 0.855 | -0.022 | 0.177 | 0.156 | -0.409 | 0.350 | 0 | 0 | 0 |
| TW | CAR-TW | Before | 0.335 | 0.332 | 0.553 | 0.708 | 0.815 | -0.010 | 0.231 | 0.083 | -0.336 | 0.558 | 0 | 0 | 0 |
| | | After | 0.407 | 0.402 | 0.543 | 0.697 | 0.770 | 0.000 | 0.317 | 0.074 | -0.292 | 0.990 | 0 | 0.0001 | 0 |
| | | Before | 0.227 | 0.223 | 0.593 | 0.794 | 0.880 | -0.015 | 0.240 | 0.083 | -0.340 | 0.364 | 0 | 0 | 0 |
| | | After | 0.321 | 0.315 | 0.496 | 0.677 | 0.824 | -0.009 | 0.321 | 0.088 | -0.305 | 0.609 | 0 | 0 | 0 |
| | TW-TW | Before | 0.327 | 0.325 | 0.631 | 0.832 | 0.821 | 0.020 | 0.220 | 0.014 | -0.369 | 0.038 | 0 | 0.1497 | 0 |
| | | After | 0.367 | 0.365 | 0.556 | 0.734 | 0.796 | 0.015 | 0.258 | 0.000 | -0.363 | 0.244 | 0 | 0.9839 | 0 |
| | | Before | 0.255 | 0.253 | 0.623 | 0.821 | 0.863 | 0.013 | 0.222 | 0.015 | -0.391 | 0.192 | 0 | 0.1408 | 0 |
| | | After | 0.309 | 0.306 | 0.581 | 0.755 | 0.832 | 0.025 | 0.275 | -0.007 | -0.354 | 0.049 | 0 | 0.5602 | 0 |
| | | Before | 0.387 | 0.372 | 0.688 | 0.923 | 0.783 | -0.035 | 0.346 | 0.123 | -0.280 | 0.339 | 0 | 0.0011 | 0 |

|  |  |  |  |  |  |  |  |  |  |  |  |  |  |  |
|---|---|---|---|---|---|---|---|---|---|---|---|---|---|---|
|  | HV-TW | After | 0.433 | 0.410 | 0.471 | 0.622 | 0.753 | -0.064 | 0.299 | 0.204 | -0.388 | 0.152 | 0 | 0 | 0 |
|  |  | Before | 0.258 | 0.239 | 0.639 | 0.806 | 0.862 | -0.008 | 0.368 | 0.171 | -0.302 | 0.825 | 0 | 0 | 0 |
|  |  | After | 0.251 | 0.220 | 0.632 | 0.818 | 0.866 | -0.008 | 0.331 | 0.218 | -0.358 | 0.852 | 0 | 0 | 0 |
|  | AUTO-TW | Before | 0.403 | 0.400 | 0.599 | 0.807 | 0.773 | -0.014 | 0.274 | 0.120 | -0.401 | 0.405 | 0 | 0 | 0 |
|  |  | After | 0.475 | 0.470 | 0.523 | 0.693 | 0.725 | -0.016 | 0.329 | 0.095 | -0.322 | 0.457 | 0 | 0 | 0 |
|  |  | Before | 0.308 | 0.304 | 0.614 | 0.834 | 0.832 | -0.007 | 0.272 | 0.092 | -0.416 | 0.680 | 0 | 0 | 0 |
|  |  | After | 0.276 | 0.269 | 0.541 | 0.735 | 0.851 | -0.019 | 0.356 | 0.091 | -0.330 | 0.354 | 0 | 0 | 0 |
| AUTO | CAR-AUTO | Before | 0.375 | 0.369 | 0.542 | 0.726 | 0.791 | -0.008 | 0.197 | 0.128 | -0.419 | 0.744 | 0 | 0 | 0 |
|  |  | After | 0.442 | 0.431 | 0.523 | 0.727 | 0.747 | -0.021 | 0.316 | 0.090 | -0.294 | 0.475 | 0 | 0.0031 | 0 |
|  |  | Before | 0.320 | 0.313 | 0.601 | 0.792 | 0.825 | -0.007 | 0.235 | 0.112 | -0.386 | 0.739 | 0 | 0 | 0 |
|  |  | After | 0.246 | 0.231 | 0.562 | 0.736 | 0.868 | -0.021 | 0.325 | 0.113 | -0.330 | 0.477 | 0 | 0.0003 | 0 |
|  | TW-AUTO | Before | 0.361 | 0.357 | 0.580 | 0.752 | 0.800 | 0.023 | 0.152 | 0.001 | -0.427 | 0.194 | 0 | 0.954 | 0 |
|  |  | After | 0.394 | 0.386 | 0.533 | 0.695 | 0.778 | 0.004 | 0.242 | -0.024 | -0.372 | 0.875 | 0 | 0.323 | 0 |
|  |  | Before | 0.200 | 0.195 | 0.622 | 0.859 | 0.895 | 0.014 | 0.198 | 0.010 | -0.427 | 0.410 | 0 | 0.564 | 0 |
|  |  | After | 0.224 | 0.214 | 0.557 | 0.723 | 0.881 | 0.026 | 0.228 | -0.027 | -0.410 | 0.273 | 0 | 0.256 | 0 |
|  | AUTO-AUTO | Before | 0.409 | 0.400 | 0.574 | 0.771 | 0.769 | 0.021 | 0.199 | -0.049 | -0.409 | 0.474 | 0 | 0.087 | 0 |
|  |  | After | 0.443 | 0.426 | 0.487 | 0.671 | 0.747 | 0.098 | 0.265 | 0.000 | -0.412 | 0.012 | 0 | 0.995 | 0 |
|  |  | Before | 0.275 | 0.264 | 0.626 | 0.841 | 0.852 | 0.046 | 0.186 | -0.080 | -0.452 | 0.106 | 0 | 0.005 | 0 |
|  |  | After | 0.292 | 0.270 | 0.602 | 0.807 | 0.842 | 0.048 | 0.286 | -0.007 | -0.387 | 0.197 | 0 | 0.859 | 0 |
| HV | CAR-HV | Before | 0.507 | 0.486 | 0.502 | 0.593 | 0.702 | 0.012 | 0.183 | 0.098 | -0.411 | 0.777 | 0.0001 | 0.022 | 0 |
|  |  | After | 0.494 | 0.452 | 0.456 | 0.629 | 0.711 | 0.049 | 0.285 | 0.198 | -0.369 | 0.342 | 0 | 0.000 | 0 |
|  |  | Before | 0.170 | 0.136 | 0.582 | 0.753 | 0.911 | -0.003 | 0.201 | 0.130 | -0.444 | 0.946 | 0 | 0.001 | 0 |
|  |  | After | 0.160 | 0.090 | 0.497 | 0.681 | 0.916 | -0.007 | 0.339 | 0.181 | -0.406 | 0.896 | 0 | 0.001 | 0 |
|  | TW-HV | Before | 0.330 | 0.315 | 0.554 | 0.723 | 0.818 | 0.009 | 0.188 | 0.064 | -0.319 | 0.803 | 0 | 0.075 | 0 |
|  |  | After | 0.426 | 0.400 | 0.400 | 0.546 | 0.758 | -0.002 | 0.245 | 0.032 | -0.340 | 0.960 | 0 | 0.458 | 0 |
|  |  | Before | 0.165 | 0.146 | 0.559 | 0.749 | 0.914 | 0.049 | 0.234 | 0.101 | -0.328 | 0.155 | 0 | 0.005 | 0 |
|  |  | After | 0.271 | 0.237 | 0.553 | 0.772 | 0.854 | 0.006 | 0.256 | 0.047 | -0.316 | 0.877 | 0 | 0.224 | 0 |

**Performance evaluation for various LV-SV class combinations**

This study proposed a three-step filtering approach: (i) filtering based on the speed–gap relationship, (ii) filtering based on approaching or diverging behaviour, and (iii) filtering based on speed correlation. The figures below illustrate the proportion and number of points removed in each of these iterations. Iteration 1 involved the most comprehensive set of filtering conditions and, consequently, removed the largest number of outlier points. Comparing the proportions of outliers removed in Iterations 2 and 3 provides insights into the behavioural differences across various leader–follower (LF) pair combinations.

For instance, in the case of CARs (refer to Figure 11 ), TW–CAR pairs exhibited the highest number of outliers in Iteration 2, followed by AUTO–CAR. This indicates that negative asymmetric pairs (i.e., those with a smaller leader vehicle and a larger follower vehicle) are more prone to diverging or approaching behaviours. A plausible explanation is that smaller leader vehicles (e.g., TWs or AUTOs) possess greater manoeuvrability, allowing them to move away more freely from the follower. In contrast, the CAR as the follower (SV) may be constrained by space, resulting in slower movement and increased outlier behaviour.

Conversely, HV–CAR pairs showed the lowest number of outliers under this criterion. This may be due to the limited manoeuvrability of heavy vehicles, which reduces erratic divergence and approaching behaviour. A similar trend is observed in Iteration 3 (based on speed correlation). When the heavy vehicle serves as the leader, the follower tends to exhibit more stable following behaviour, resulting in fewer outliers. This could be attributed to reduced visibility and manoeuvring space behind a heavy vehicle, which may discourage lateral movements and lead to a stronger speed correlation. Therefore, positive asymmetric pairs such as HV–CAR demonstrate more consistent and stable car-following behaviour.

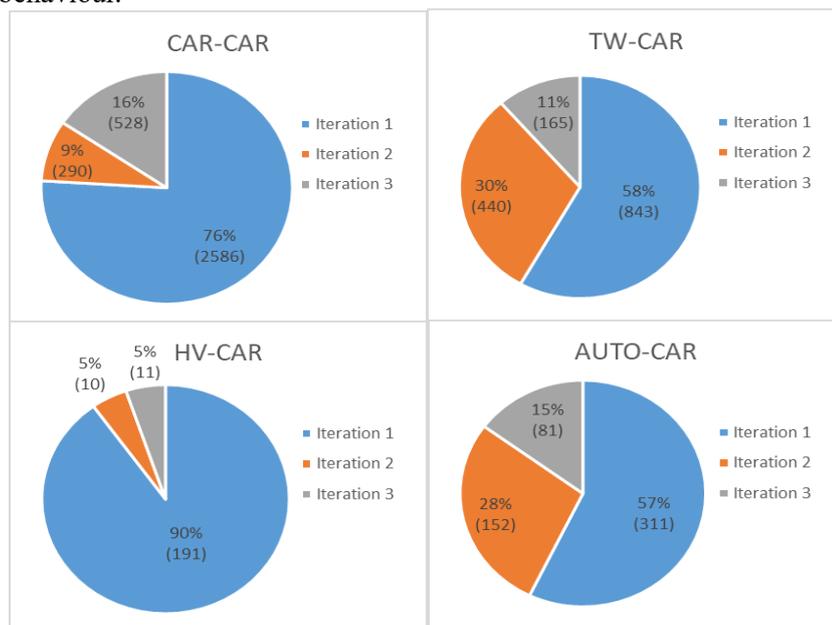

**Figure 11 Proportion of removed points for CAR**

For subject vehicle (SV) type TW, a similar trend is observed (refer to Figure 12). TWs are involved in only two categories of LF pairings: symmetric (TW–TW) and positive asymmetric (e.g., CAR–TW, AUTO–TW, HV–TW). Among these, the symmetric TW–TW pairs exhibit the highest number of outliers in both Iteration 2 and Iteration 3. Due to their smaller size and high manoeuvrability, TWs often demonstrate irregular movement patterns, making it difficult to maintain consistent vehicle-following behaviour. In contrast, when the leader vehicle is larger, TWs tend to show improved following behaviour, likely due to constrained manoeuvrability and reduced opportunities for lateral movement, resulting in fewer outliers.

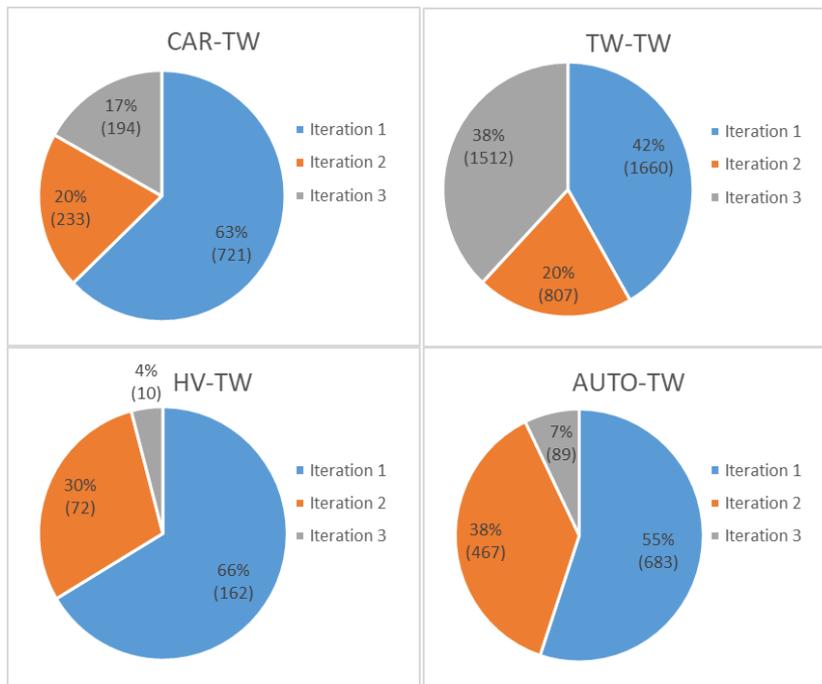

**Figure 12 Proportion of removed points for TW**

For the SV type AUTO, all three combinations—CAR–AUTO, TW–AUTO, and AUTO–AUTO—exhibited a similar proportion of outlier points in Iteration 2. This may be attributed to the AUTO's small size and high manoeuvrability, which enable more dynamic movement. Additionally, as AUTOs typically operate as taxis, their drivers may exhibit more aggressive driving behaviour to minimise travel time and maximise the number of trips, thereby increasing income. In Iteration 3, both positive and negative asymmetric AUTO-related pairs followed trends similar to those observed in other LF combinations, showing comparable speed correlation characteristics (refer to Figure 13).

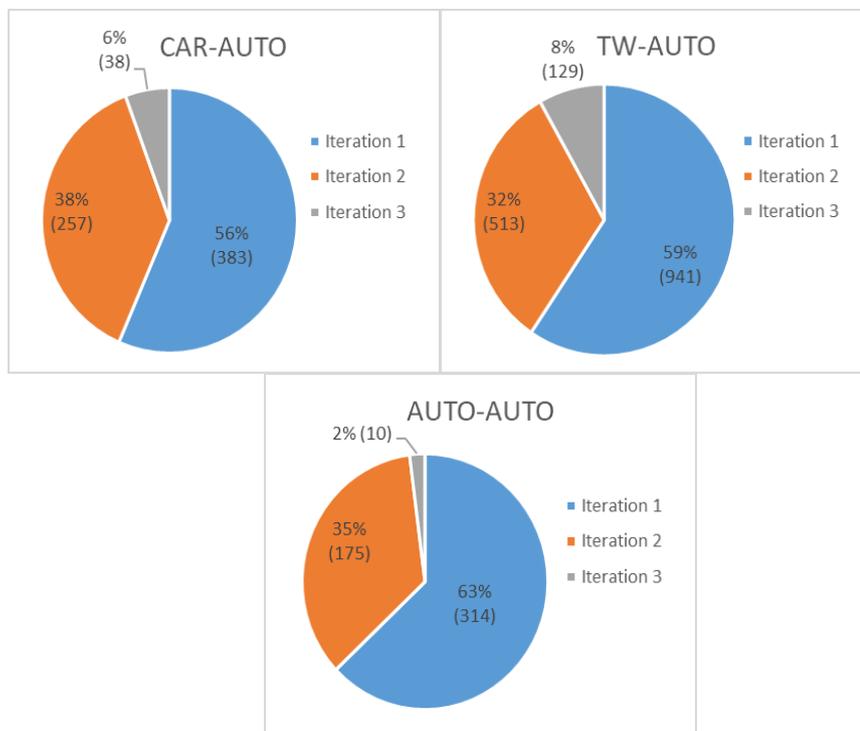

**Figure 13 Proportion of removed points for AUTO**

For the HV category, no outlier points were identified in Iteration 3, indicating strong speed correlation in HV-related pairs. As expected, TW–HV pairs exhibited a higher proportion of outliers in Iteration 2 compared to CAR–HV pairs, likely due to the greater behavioral differences and maneuverability contrast between TWs and HVs (refer to Figure 14).

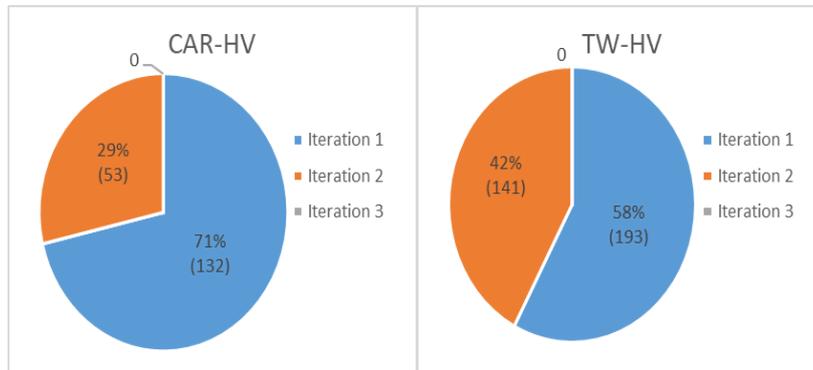

**Figure 14 Proportion of removed points for HV**

**Weight distribution**

In addition to traditional performance metrics, the current study also evaluated the **b** as an additional diagnostic parameter to evaluate the proposed methodology. Figure 16 illustrates the weight distributions for both retained and removed data points across different LF pair combinations (see Figure 15 for a representative example of CAR-CAR).

The figures indicate that the proposed filtering methodology effectively eliminated data points associated with lower weights, which typically contribute more to model error. In contrast, the retained points generally exhibit higher weights and are associated with smaller residuals, suggesting that these points better represent the underlying vehicle-following behaviour and contribute more reliably to the model's accuracy.

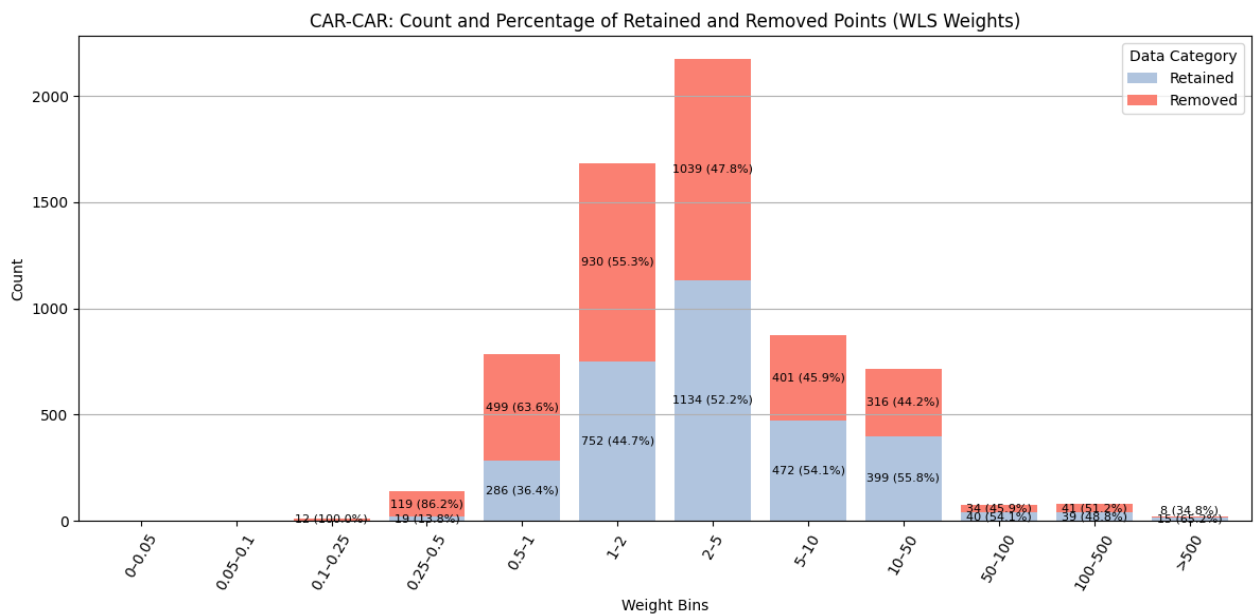

**Figure 15 CAR-CAR weights distribution**

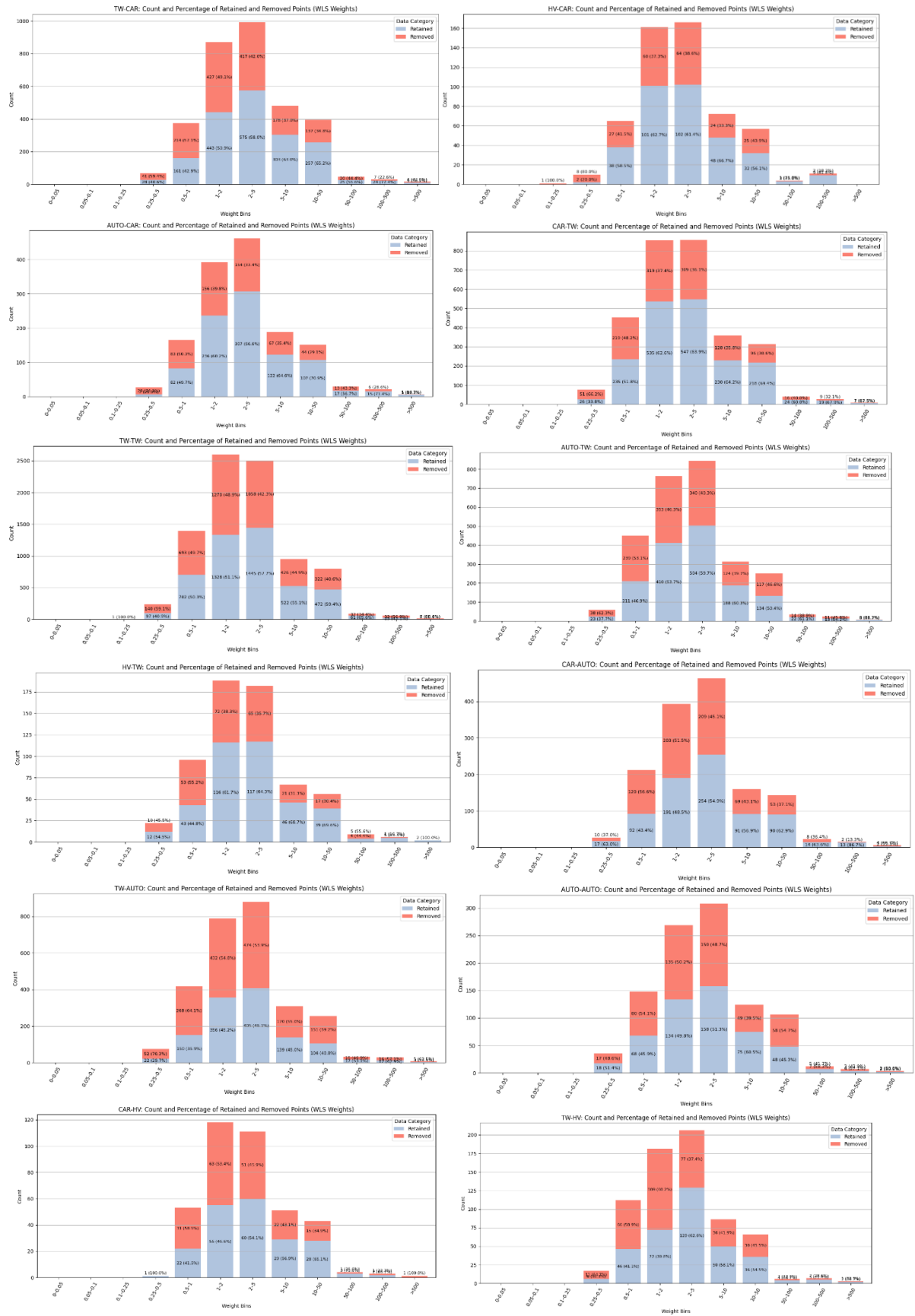

**Figure 16 Weight distribution of retained and removed points across LF pairs**

Incorporating vehicle-type-specific longitudinal gaps and speed profile correlations through wavelet energy analysis significantly improved LF pair identification and model accuracy. Instead of routinely selecting the pairs within a longitudinal gap of 30m, this methodology cross-verified the pairs based on speeds and gaps. Figure 17 gives the details of the longitudinal gap distribution of both retained and removed data points. The methodology accurately removed pairs, maintaining very low and very large longitudinal gaps. Similarly, Figure 18 and Figure 19 gives the distribution of longitudinal gap range, which demonstrates the removal of pairs maintaining larger longitudinal gap ranges and distribution of relative velocity sign change ratio, respectively. Those are the pairs that exhibit an approaching or diverging behaviour instead of proper car-following behaviour. The threshold of 10m selected for the methodology was found to be accurate from the graph, since the majority of pairs maintaining a larger gap range got removed by this methodology. Figure 20 gives details of the relative velocity distribution. The chosen threshold value of ±2.5 m/s appears to be a suitable criterion since 95 percentile of retained values comes within this range.

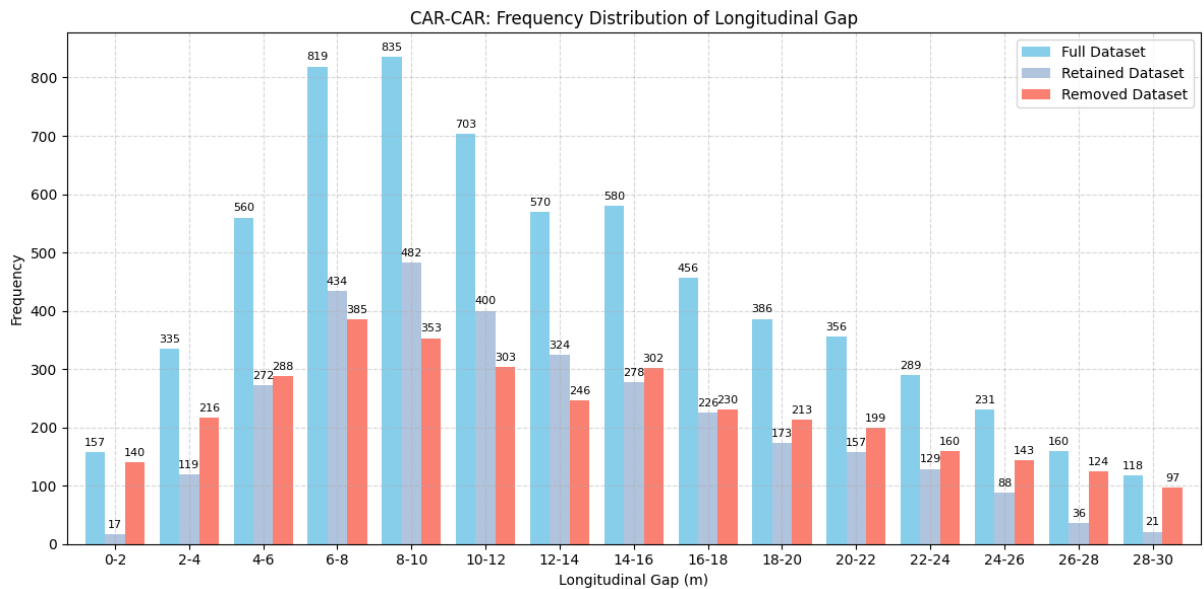

**Figure 17 CAR-CAR longitudinal gap distribution**

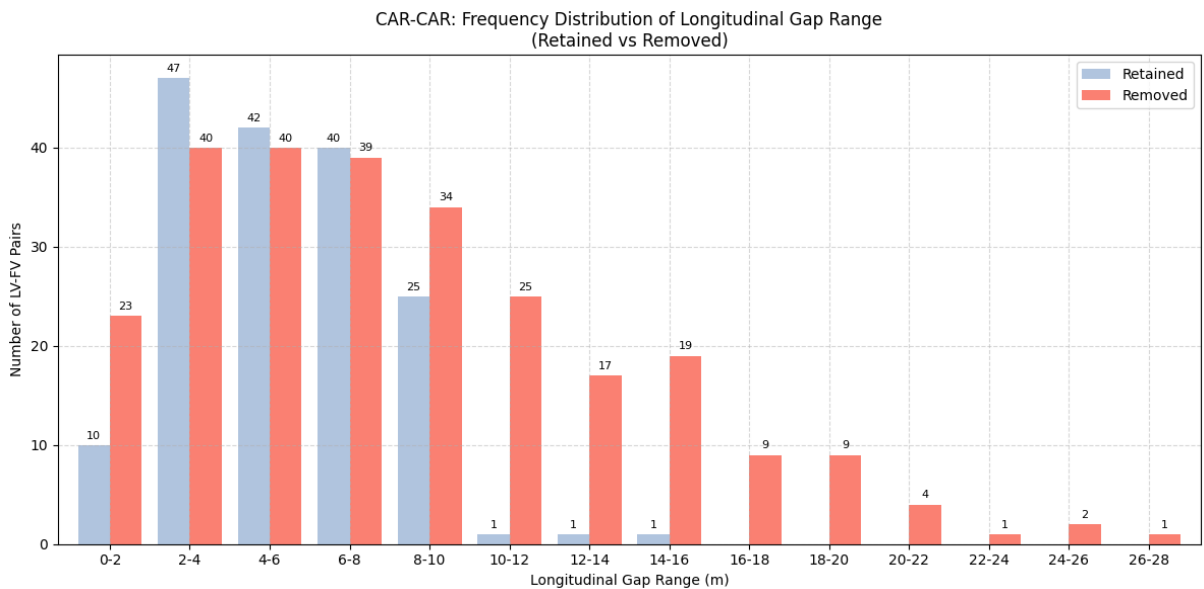

**Figure 18 CAR-CAR longitudinal gap range distribution**

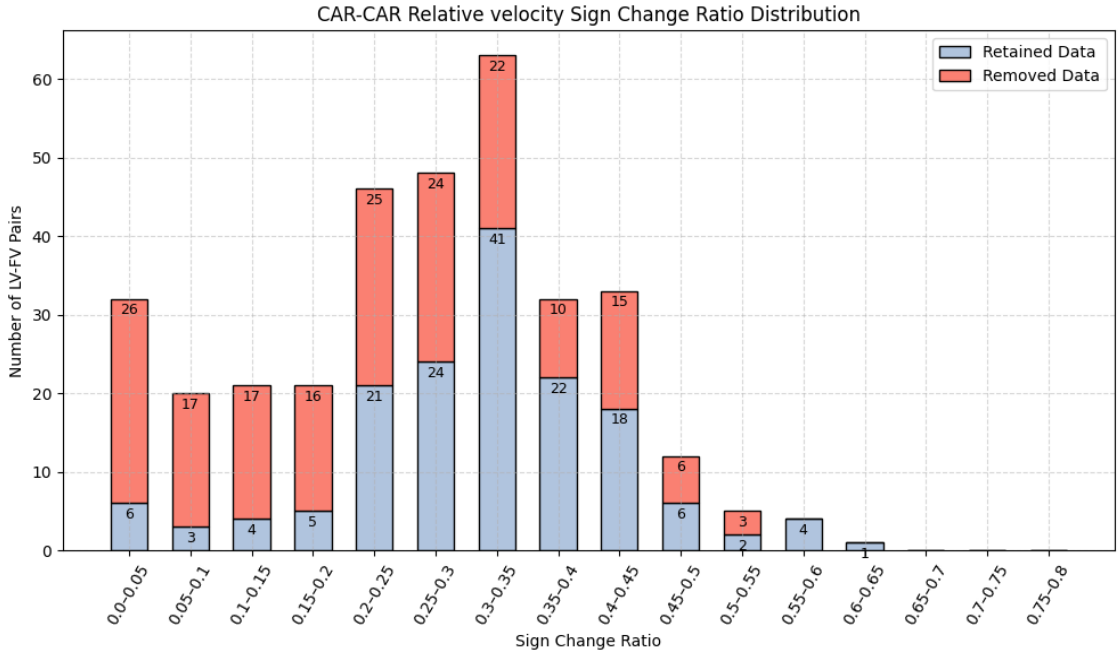

**Figure 19 CAR-CAR relative velocity sign change ratio**

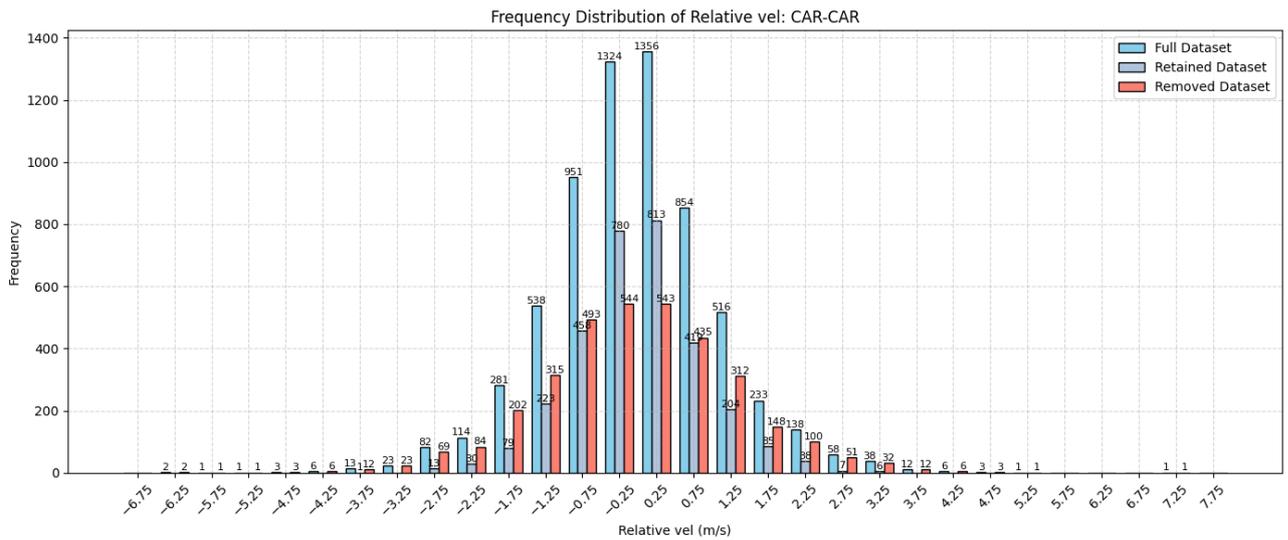

**Figure 20 CAR-CAR relative velocity distribution**

**CONCLUSION**

This study presents a novel methodology for identifying valid LF pairs under HWLD traffic conditions. Initially, LF pairs were selected based on criteria from the literature, which showed poor correlation between longitudinal gap and speed. Further investigation revealed three distinct behavioural patterns indicating the presence of non-following behaviour within these LF pairs:

i. The SV maintains very small longitudinal gaps at high speeds.
ii. The SV followed at large longitudinal gaps while moving at low speeds—likely reflecting weak influence from distant LVs. This behaviour is more pronounced for TWs, whose look-ahead distance is typically shorter than that of HVs.
iii. The longitudinal gap changed significantly over time (either increasing or decreasing), without corresponding changes in SV speed, suggesting approaching or diverging motion rather than consistent following behaviour.

These patterns indicate that physical positioning alone - LV within 30m with lateral overlap for a minimum following duration (5s) - is not sufficient to confirm VF behaviour. In many cases, the SV was not influenced by the LV, though they satisfied the positional requirements, underscoring the need for filtering non-following pairs to improve the reliability of VF models.

To address this, the study leveraged the q-k fundamental diagram to estimate appropriate longitudinal gaps based on vehicle type and speed. Additionally, a Mexican Hat Wavelet Transform (MWT) is used to identify potential LVs that show a meaningful influence on the SV. Based on these insights, a three-phase filtering methodology was proposed, comprising:

  i. filtering based on the speed–gap relationship
 ii. filtering based on approaching or diverging behaviour, and
iii. filtering based on speed correlation.

This multi-stage filtering approach effectively eliminated pairs exhibiting overtaking initiation, tailgating, and inconsistent gap dynamics. The first phase removed pairs with undesirable gap-speed combinations, typically representing the overtaking initiation or tailgating (76% for CAR-CAR). The second phase excluded pairs with poor relative velocity trends, particularly those with longitudinal gaps greater than 10 meters and a low relative velocity sign change ratio (<0.3), indicating approaching or diverging behaviour (9% for CAR-CAR). The third phase applied wavelet-based speed correlation to retain only those pairs where the LV influenced SV movement (16% for CAR-CAR).

Through this rigorous filtering, only those LF pairs (167 pairs out of 363 for CAR-CAR) exhibiting consistent and coherent VF behaviour were retained, significantly enhancing the robustness and interpretability of the subsequent modelling. On this refined dataset, the study further explored behavioural differences across various LF pair combinations. The analysis confirmed that both LV type and SV type significantly influence vehicle-following behaviour. Based on vehicle size asymmetry, pairs were categorized into symmetric, positive asymmetric, and negative asymmetric groups. While similar VF trends were generally observed within each group, certain combinations deviated from this pattern, highlighting the complexity of interactions in heterogeneous traffic.

The major contributions of this study are as follows:

   i. The proposed filtering methodology allows for adaptable threshold calibration depending on the combination of LV–SV. This flexibility enables tailored identification of valid LF interactions across heterogeneous traffic conditions.
  ii. The study introduces innovative plots combining multiple variables to comprehensively examine LF behaviour. These visual tools provide deeper insights into vehicle-following patterns across different traffic scenarios.
 iii. Several novel metrics—such as *maximum gap range* and *sign change ratio*—are proposed for identifying potential outliers within LF pairs. These metrics enhance the precision of the filtering process by capturing subtle inconsistencies in LF interactions.
  iv. The gap-based filtering approach, which incorporates equilibrium spacing principles, removed up to 38% of LF pairs and 76% of data points in the CAR–CAR case. This result underscores the importance of using realistic spacing models rather than fixed distance thresholds in identifying valid LF pairs.
   v. Unlike traditional methods, which overlook dynamic interactions such as approaching or diverging motion, the proposed methodology effectively captures such behaviours. In the CAR–CAR case, this filter eliminated 8% of pairs and 9% of data points, demonstrating its effectiveness in refining LF identification.
  vi. Traditional proximity-based methods fail to verify whether the LV genuinely influences the SV's speed profile. The study addresses this gap by employing the MWT to detect delayed SV responses relative to the LV. This filter removed 19% of pairs and 16% of points in the CAR–CAR scenario, providing a behaviourally grounded filtering layer.
 vii. The filtering methodology led to substantial improvements in model accuracy across all 13 LF pair combinations. For instance, in the TW–HV combination, the $R^2$ value improved by 49.3%, with 48.83% of the data removed through filtering. The corresponding reductions in MAE,

|      | RMSE, and NRMSE were 18%, 23%, and 5.6%, respectively. Even in less responsive pairs like HV–CAR, improvements were observed, with a 3% increase in $R^2$ and respective MAE, RMSE, and NRMSE reductions of 7.8%, 13%, and 0.7%. |
|------|---|
| viii. | The filtered points consistently exhibited low regression weights (predominantly in the 0–5 range) when applying a weighted regression model, validating their exclusion. Additionally, for the CAR–CAR combination, the longitudinal gap emerged as a significant predictor only after filtering, reinforcing the importance of isolating valid vehicle-following behaviour. |
| ix. | Based on a detailed empirical analysis, the study proposes a systematic and transferable filtering algorithm that can be applied to other datasets to improve LF pair identification in diverse traffic environments. |
| x. | Unlike black-box techniques (weighted regression), the proposed methods are grounded in established traffic flow principles, making them interpretable, transparent, and practical for both research and application in traffic modelling and simulation. |

While this study offers valuable insights, it has certain limitations. The analysis focused only on overlapping LF pairs, excluding non-overlapping interactions. Desired gap estimation was based on subject vehicle-specific fundamental diagrams, which may not fully account for behavioural variability across LV types. The threshold-based filtering approach also requires separate calibration for each LF combination, which can be computationally intensive. Additionally, the findings are based on data from a single location, limiting generalisability. Future work can address these limitations by incorporating non-overlapping pairs, adopting adaptive or vehicle-type-specific thresholds, integrating behavioural variability into gap estimation, and validating the approach across diverse urban settings.

Overall, this comparative analysis highlights the significant variability in VF behaviour across different LF pair types and underscores the importance of context-specific filtering when modelling traffic dynamics in HWLD traffic. The proposed framework provides a robust foundation for accurately extracting meaningful LF pairs, with promising applications in traffic flow modelling, safety assessment, and driver behaviour analysis.

**ACKNOWLEDGMENTS**

The authors are thankful to Kanagaraj et al. (2015) for open access to the mixed traffic trajectory data. During the preparation of this work, the author(s) used ChatGPT and Grammarly to do grammar checks and language refinement. All content was subsequently reviewed and edited by the authors, who take full responsibility for the final version of the manuscript.

**AUTHOR CONTRIBUTIONS**

The authors confirm contribution to the paper as follows: study conception and design: Eldhose, Chilukuri, Rajendran; data collection: not applicable; analysis and interpretation of results: Eldhose, Chilukuri, Rajendran; draft manuscript preparation: Eldhose, Chilukuri, Rajendran. All authors reviewed the results and approved the final version of the manuscript.

**REFERENCES**

1. Madhu, K., K. K. Srinivasan, and R. Sivanandan. Following Behaviour in Mixed Traffic: Effects of Vehicular Interactions, Local Area Concentration and Driving Regimes. *International Journal of Engineering Research and Technology*, Vol. 13, No. 6, 2020, pp. 1353–1368. https://doi.org/10.37624/ijert/13.6.2020.1353-1368.

2. Mathew, T. V., and K. V. R. Ravishankar. Car-Following Behavior in Traffic Having Mixed Vehicle-Types. *Transportation Letters*, Vol. 3, No. 2, 2011, pp. 109–122. https://doi.org/10.3328/TL.2011.03.02.109-122.

3. Das, S., A. K. Maurya, and A. K. Budhkar. Determinants of Time Headway in Staggered Car-Following Conditions. *Transportation Letters*, Vol. 11, No. 8, 2019, pp. 447–457. https://doi.org/10.1080/19427867.2017.1386872.

4. Mathew, T. V., and P. Radhakrishnan. Calibration of Microsimulation Models for Nonlane-Based Heterogeneous Traffic at Signalized Intersections. *Journal of Urban Planning and Development*, Vol. 136, No. 1, 2010, pp. 59–66. https://doi.org/10.1061/(asce)0733-9488(2010)136:1(59).

5. Asaithambi, G., V. Kanagaraj, K. K. Srinivasan, and R. Sivanandan. Study of Traffic Flow Characteristics Using Different Vehicle-Following Models under Mixed Traffic Conditions. *Transportation Letters*, Vol. 10, No. 2, 2018, pp. 92–103. https://doi.org/10.1080/19427867.2016.1190887.

6. Madhu, K., K. K. Srinivasan, and R. Sivanandan. Acceleration Models for Two-Wheelers and Cars in Mixed Traffic: Effect of Unique Vehicle-Following Interactions and Driving Regimes. *Current Science*, Vol. 122, No. 12, 2022, pp. 1441–1450. https://doi.org/10.18520/cs/v122/i12/1441-1450.

7. Nirmale, S. K., A. R. Pinjari, and A. Sharma. A Discrete-Continuous Multi-Vehicle Anticipation Model of Driving Behaviour in Heterogeneous Disordered Traffic Conditions. *Transportation Research Part C: Emerging Technologies*, Vol. 128, No. November 2020, 2021, p. 103144. https://doi.org/10.1016/j.trc.2021.103144.

8. Anand, P. A., P. Atmakuri, V. S. R. Anne, G. Asaithambi, K. K. Srinivasan, R. Sivanandan, and B. R. Chilukuri. Calibration of Vehicle-Following Model Parameters Using Mixed Traffic Trajectory Data. *Transportation in Developing Economies*, Vol. 5, No. 2, 2019, pp. 1–11. https://doi.org/10.1007/s40890-019-0086-4.

9. Raju, N., S. Arkatkar, and G. Joshi. Modeling Following Behavior of Vehicles Using Trajectory Data under Mixed Traffic Conditions: An Indian Viewpoint. *Transportation Letters*, Vol. 13, No. 9, 2021, pp. 649–663. https://doi.org/10.1080/19427867.2020.1751440.

10. Papathanasopoulou, V., and C. Antoniou. Flexible Car – Following Models for Mixed Traffic and Weak Lane – Discipline Conditions. *European Transport Research Review*, Vol. 10, 2018, pp. 1–14.

11. Kulkarni, M. M., A. A. Chaudhari, K. K. Srinivasan, B. R. Chilukuri, M. Treiber, and O. Okhrin. Leader–Follower Identification with Vehicle-Following Calibration for Non-Lane-Based Traffic. *Transportation Research Part C: Emerging Technologies*, Vol. 171, No. December 2024, 2025, p. 104940. https://doi.org/10.1016/j.trc.2024.104940.

12. Maiti, N., and B. R. Chilukuri. Empirical Investigation of Fundamental Diagrams in Mixed Traffic. *IEEE Access*, Vol. 11, No. January, 2023, pp. 13293–13308. https://doi.org/10.1109/ACCESS.2023.3242971.

13. Kanagaraj, V., G. Asaithambi, T. Toledo, and T. C. Lee. Trajectory Data and Flow Characteristics of Mixed Traffic. *Transportation Research Record*, Vol. 2491, 2015, pp. 1–11.

https://doi.org/10.3141/2491-01.

14. Kulkarni, M. M., A. A. Chaudhari, K. Karthik, B. R. Chilukuri, M. Treiber, and O. Okhrin. Leader-Follower Identification with Vehicle-Following Calibration for Non-Lane-Based Traffic. *arXiv preprint arXiv:2405.10665*, 2024, pp. 1–45.

15. Chaudhari, A. A., K. K. Srinivasan, B. R. Chilukuri, M. Treiber, and O. Okhrin. CalibratingWiedemann-99 Model Parameters to Trajectory Data of Mixed Vehicular Traffic. *Transportation Research Record*, Vol. 2676, No. 1, 2022, pp. 718–735. https://doi.org/10.1177/03611981211037543.

16. Ashok, A., and B. R. Chilukuri. A Framework to Characterise, Estimate, and Predict Vehicle Class-Agnostic Traffic States and Class-Wise Speeds for Mixed Traffic Conditions. *IEEE Access*, Vol. 12, No. August, 2024, pp. 106211–106235. https://doi.org/10.1109/ACCESS.2024.3437172.

17. Raju, N., S. Arkatkar, and G. Joshi. Evaluating Performance of Selected Vehicle Following Models Using Trajectory Data under Mixed Traffic Conditions. *Journal of Intelligent Transportation Systems: Technology, Planning, and Operations*, Vol. 24, No. 6, 2020, pp. 617–634. https://doi.org/10.1080/15472450.2019.1675522.

18. Kashyap, N. M., B. R. Chilukuri, K. K. Srinivasan, and G. Asaithambi. Analysis of Vehicle-Following Behavior in Mixed Traffic Conditions Using Vehicle Trajectory Data. *Transportation Research Record*, Vol. 2674, No. 11, 2020, pp. 842–855. https://doi.org/10.1177/0361198120949874.